\input harvmac.tex
\input epsf.sty


\lref\coleman{
  S.~R.~Coleman,
  ``The Fate Of The False Vacuum. 1. Semiclassical Theory,''
  Phys.\ Rev.\ D {\bf 15}, 2929 (1977)
  [Erratum-ibid.\ D {\bf 16}, 1248 (1977)].
  C.~G.~.~Callan and S.~R.~Coleman,
  ``The Fate Of The False Vacuum. 2. First Quantum Corrections,''
  Phys.\ Rev.\ D {\bf 16}, 1762 (1977).
}
 .


%

{\Title{\vbox{
\hbox{HUTP/A040}
}}
{\vbox{
\centerline{Geometrically Induced Metastability}
\centerline{and}
\centerline{Holography}}}
\vskip .3in
\centerline{Mina Aganagic$^1$, Christopher Beem$^1$, Jihye Seo$^2$, and Cumrun
Vafa$^{2,3}$}
\vskip .4in
}

\centerline{$^1$ University of California, Berkeley, CA 94720}
\centerline{$^2$ Jefferson Physical Laboratory, Harvard University,
Cambridge, MA 02138}
\centerline{$^3$ Center for Theoretical Physics, Massachusetts Institute of Technology,
Cambridge, MA 02139}

\vskip .4in

We construct metastable configurations of branes and anti-branes
wrapping 2-spheres inside local Calabi-Yau manifolds
and study their large $N$ duals.  These duals are Calabi-Yau
manifolds in which the wrapped 2-spheres have been replaced by
3-spheres with flux through them, and
supersymmetry is spontaneously broken.  The geometry of the
non-supersymmetric vacuum is exactly calculable to all orders of
the `t Hooft parameter, and to the leading order in $1/N$.  The
computation utilizes the same matrix model techniques that were
used in the supersymmetric context.  This provides
a novel mechanism for breaking supersymmetry in the context
of flux compactifications.

\vfill
\eject

\newsec{Introduction}

One of the central questions currently facing string theory is how to break
supersymmetry in a controllable way.  The most obvious ways to break it typically
lead to instabilities signaled by the appearance of tachyons
in the theory.  One would like to find vacua in which
supersymmetry is broken, but stability is not lost.  It seems
difficult (or impossible, at present) to obtain exactly stable
non-supersymmetric vacua from string theory.  Therefore, the only candidates
would appear to be meta-stable non-supersymmetric vacua.  This idea has already been realized
in certain models (See \ref\meta{ M.~R.~Douglas and S.~Kachru,
  ``Flux compactification,''
  arXiv:hep-th/0610102.
 }\ for a review
and the relevant literature). More recently, the fact that metastable vacua are also generic
in ordinary supersymmetric gauge theories \ref\sis{
  K.~Intriligator, N.~Seiberg and D.~Shih,
  ``Dynamical SUSY breaking in meta-stable vacua,''
  JHEP {\bf 0604}, 021 (2006)
  [arXiv:hep-th/0602239].
}\
has added further motivation
for taking this method of breaking supersymmetry 
seriously within string theory.
Potential realizations of such metastable gauge theories
have been considered in string
theory \ref\oth{H.~Ooguri and Y.~Ookouchi,
  ``Meta-stable supersymmetry breaking vacua on intersecting branes,''
  arXiv:hep-th/0607183;
  S.~Franco, I.~Garcia-Etxebarria and A.~M.~Uranga,
  ``Non-supersymmetric meta-stable vacua from brane configurations,''
  arXiv:hep-th/0607218;
    I.~Bena, E.~Gorbatov, S.~Hellerman, N.~Seiberg and D.~Shih,
  ``A note on (meta)stable brane configurations in MQCD,''
  arXiv:hep-th/0608157.
} (see also \ref\othnonsu{
  S.~Kachru and J.~McGreevy,
  ``Supersymmetric three-cycles and (super)symmetry breaking,''
  Phys.\ Rev.\ D {\bf 61}, 026001 (2000)
  [arXiv:hep-th/9908135];

      K.~A.~Intriligator, N.~Seiberg and S.~H.~Shenker,
  ``Proposal for a simple model of dynamical SUSY breaking,''
  Phys.\ Lett.\ B {\bf 342}, 152 (1995)
  [arXiv:hep-ph/9410203].
}).

The aim of this paper is to study an alternative approach to breaking
supersymmetry via metastable configurations, as suggested
in \ref\vafa{
  C.~Vafa,
   ``Superstrings and topological strings at large N,''
  J.\ Math.\ Phys.\  {\bf 42}, 2798 (2001)
  [arXiv:hep-th/0008142].
}.  In this scenario, we wrap branes and anti-branes on
cycles of local Calabi-Yau manifolds, and metastability
is a consequence of the Calabi-Yau geometry.
In a sense, this is a geometrically induced
metastability.  The branes and the anti-branes are wrapped over
2-cycles which are rigid and separated.  In order for the branes
to annihilate, they have to move, which costs energy
as the relevant minimal 2-spheres are rigid.  This leads to a potential
barrier due to the stretching of the brane and results in a
configuration which is metastable.  It is particularly interesting
to study the same system at large $N$, where we have a large
number of branes and anti-branes.  In this case, it is better
to use a dual description obtained via a geometric transition
in which the 2-spheres are shrunk, and get replaced by 3-spheres with
fluxes through them.  The dual theory has ${\cal N}=2$
supersymmetry, which the flux breaks spontaneously. If we have
only branes in the original description, then the supersymmetry is broken to
an ${\cal N}=1$ subgroup. With only anti-branes present, we expect it to be
broken to a {\it different}
${\cal N}=1$ subgroup, and with both branes {\it and} anti-branes, the supersymmetry
should be completely broken. The vacuum structure can be analyzed
from a potential which can be computed exactly using topological string theory
\ref\civ{
  F.~Cachazo, K.~A.~Intriligator and C.~Vafa,
  ``A large N duality via a geometric transition,''
  Nucl.\ Phys.\ B {\bf 603}, 3 (2001)
  [arXiv:hep-th/0103067].
}\ or matrix models \ref\dv{
  R.~Dijkgraaf and C.~Vafa,
  ``A perturbative window into non-perturbative physics,''
  arXiv:hep-th/0208048.
}.

Unlike the cases studied before -- involving only branes --
with branes and anti-branes present, we expect to find
a meta-stable vacuum which breaks supersymmetry.
We will find that this is the case, and moreover this leads
to a controllable way of breaking supersymmetry at large $N$
where to all orders in the `t Hooft coupling, but to leading order in the
$1/N$ expansion, we can compute the geometry of the vacua and the low
energy Lagrangian.

The organization of this paper is as follows: In section 2 we review
the case where we have a single stack of branes and extend it to the
case where we have a single stack of anti-branes.  In section 3 we
discuss the case with more than one $S^2$ in the geometry, and then specialize to
the case in which there are only two.  We'll explain how if we
have only branes or only anti-branes, supersymmetry is not broken,
whereas if we have branes wrapped on one $S^2$ and anti-branes on the other,
supersymmetry is spontaneously broken.  In section 4 we estimate the
decay rate of the metastable vacuum.  In section 5 we conclude with
open questions and suggestions for future work.

\newsec{Branes and anti-branes on the conifold}

Let us begin by recalling the physics of $N$ D5 branes on the conifold
singularity.  We consider type IIB string theory with the branes
wrapping the $S^2$ obtained by resolving the conifold
singularity.  The local geometry of the Calabi-Yau threefold is a
${\bf {P}^1}$ with normal bundle given by the sum of two line
bundles
$$O(-1)+O(-1) \rightarrow \bf {P}^1 $$
%
%
At low energies, the field theory on the space-filling D5 branes is a
pure $U(N)$ gauge theory with ${\cal N}=1$ supersymmetry.
That the theory has
${\cal N}=1$ supersymmetry follows from the fact that string
theory on the background of local Calabi-Yau manifolds preserves
${\cal N}=2$, and the D-branes break half of it.
In particular, there are no massless adjoint fields, which
follows from the important
geometric fact that the ${\bf {P}^1}$ which the D5 brane wraps is {\it isolated}
(its normal bundle is $O(-1)+O(-1)$, which does not have a holomorphic
section). In other words, any deformation of the ${\bf {P}^1}$ in the normal direction
increases its volume, and so corresponds to a
massive adjoint field, at best. (see Fig. 1)
\eject
\bigskip
\centerline{\epsfxsize 3.0truein\epsfbox{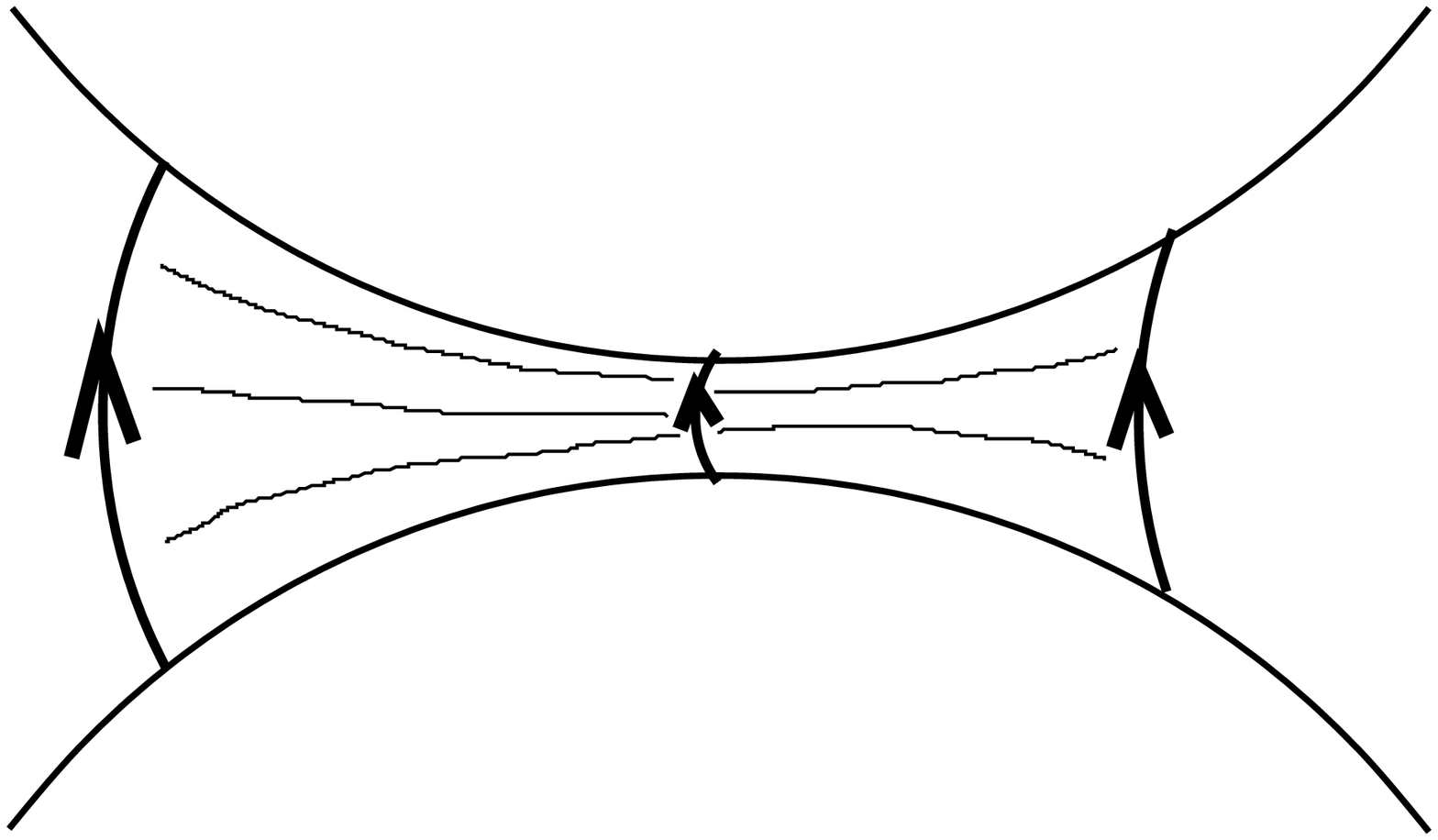}}

\noindent{\ninepoint
\baselineskip=2pt {\bf Fig. 1.} {The geometry near a resolved conifold. Moving away in the normal direction, the volume of the wrapped 2-cycle (represented here as an $S^1$) must increase.}}
\bigskip

%
%

The ${\cal N}=1$ pure Yang-Mills theory on the brane
is expected to be strongly coupled in the IR,
leading to gaugino bilinear condensation, confinement, and a mass gap.
There are $N$ massive vacua corresponding to the gaugino superfield
getting an expectation value:
\eqn\crit{ \vev{S} = \Lambda_0^{3} \;\exp({-{2 \pi i \alpha \over N}
})\exp({2\pi ik\over N})\qquad k=1,\ldots,N\; }
In the future we will suppress the phase factor which
distinguishes the $N$ vacua. Above, $\alpha$ is the
bare gauge coupling constant defined at the cutoff scale $\Lambda_0$
$$
\alpha(\Lambda_0) = -{\theta\over 2 \pi} - i {4\pi \over {g_{\rm
YM}}^2(\Lambda_0)},
$$
and
$
S={1\over 32
\pi^2} {\rm Tr} W_{\alpha} W^{\alpha}.
$
Recalling that the gauge coupling in this theory runs as
\eqn\run{
2\pi i \alpha(\Lambda_0) = -\log ({\Lambda\over \Lambda_0})^{3N}
}
where ${\Lambda}$ is the strong coupling scale of the theory,
we can also write \crit\ as
\eqn\critwo{
\vev{S} =\Lambda^{3}
}
The theory has an anomalous axial $U(1)_R$ symmetry
which rotates the gauginos according to $\lambda \rightarrow \lambda e^{i  \varphi}$, so
$$
S \rightarrow S e^{2 i \varphi}.
$$
The anomaly means that this is a symmetry of the theory only provided
the theta angle shifts
\eqn\thetaeq{
\theta \rightarrow \theta + 2 N \varphi.
}
Since the theta angle shifts, the tree level
superpotential ${\cal W}_{\rm tree} =  \alpha S$ is not invariant under the R-symmetry,
and in the quantum theory additional terms must be generated to correct for this.
  Adding the correction terms produces the effective Veneziano-Yankielowicz superpotential
\ref\VY{
  G.~Veneziano and S.~Yankielowicz,
   ``An Effective Lagrangian For The Pure N=1 Supersymmetric Yang-Mills
  Theory,''
  Phys.\ Lett.\ B {\bf 113}, 231 (1982).
},
\eqn\eff{ {\cal W}(S) = \alpha S + {1 \over 2\pi i} N S\,
(\log(S/\Lambda_0^3)-1) }
whose critical points are \crit .


%
%
%

One can also understand the generation of this superpotential from the
viewpoint of the large $N$ holographically dual theory
\vafa. This configuration of branes wrapping a
${\bf P}^1$ is dual to a closed string theory on
the Calabi-Yau manifold obtained by a geometric transition which
replaces the wrapped ${\bf P}^1$ with a finite sized
${\bf S}^3$.
%
%
In the dual theory, the 5-branes have disappeared, and in their place
there are $N$ units of Ramond-Ramond flux
$$
\oint_{A} H = N
$$
through the 3-cycle $A$ corresponding to the $S^3$.
Here
$$
H = H^{RR} + \tau H^{NS}
$$
where $\tau$ is the type IIB dilaton-axion, $\tau = C_0 + {i\over g_s}$.
There are also fluxes turned on through the dual $B$ cycle
$$
\int_{B} H =- \alpha.
$$
%

These fluxes generate a superpotential
\ref\GVW{
  S.~Gukov, C.~Vafa and E.~Witten,
  ``CFT's from Calabi-Yau four-folds,''
  Nucl.\ Phys.\ B {\bf 584}, 69 (2000)
  [Erratum-ibid.\ B {\bf 608}, 477 (2001)]
  [arXiv:hep-th/9906070].}\ref\tva{
  T.~R.~Taylor and C.~Vafa,
  ``RR flux on Calabi-Yau and partial supersymmetry breaking,''
  Phys.\ Lett.\ B {\bf 474}, 130 (2000)
  [arXiv:hep-th/9912152].

}:
$$
{\cal W} = \int
H \wedge {\Omega}
$$
where $\Omega$ is the holomorphic $(3,0)$ form on the Calabi-Yau manifold.
This can be written in terms of the periods
$$
S = \oint_{A} \Omega, \qquad
{\partial
\over \partial S} {\cal F}_0= \int_B \Omega
$$
as
\eqn\supcon{
{\cal W}(S) = \alpha \;S +N\;
{\partial \over \partial S} {\cal F}_0.
}
Above, ${\cal F}_0$ is the prepotential of the ${\cal N}=2$ $U(1)$ gauge theory which is the low-energy effective theory of type IIB string theory on this geometry before turning on fluxes.

For our case, the $B$-period was computed in \vafa\
$$
{\partial \over \partial S} {\cal F}_0= {1\over 2 \pi i} \; S\,
(\log(S/\Lambda_0^3)-1)
$$
and so \supcon\ exactly reproduces the effective superpotential \eff\ which
we derived from gauge theory arguments.
%
%
Moreover, the superpotential is an F-term which should not depend on
 the cutoff $\Lambda_0$, so
the flux has to run (this agrees with the interpretation of $\Lambda_0$ as an IR cutoff
in the conifold geometry which regulates the non-compact $B$-cycle).
$$
2\pi i \alpha(\Lambda_0) = -\log ({\Lambda\over \Lambda_0})^{3N}
$$
In the dual gauge theory, this corresponds to the running of the
gauge coupling in the low energy theory.  This theory and the duality
were studied from a different perspective in \ref\klebs{
 I.~R.~Klebanov and M.~J.~Strassler,
   ``Supergravity and a confining gauge theory: Duality cascades and
  chiSB-resolution of naked singularities,''
  JHEP {\bf 0008}, 052 (2000)
  [arXiv:hep-th/0007191].

  }\
(see also \ref\malnu{

  J.~M.~Maldacena and C.~Nunez,
  ``Towards the large N limit of pure N = 1 super Yang Mills,''
  Phys.\ Rev.\ Lett.\  {\bf 86}, 588 (2001)
  [arXiv:hep-th/0008001].}).

\subsec{The anti-brane holography}

Now consider replacing the D5 branes with anti-D5 branes
wrapping the ${\bf P^1}$.  It is natural that the physics in the
presence of the two types of branes should be identical.
In particular, in the open string theory, we again expect gaugino
condensation, a mass gap and confinement.
We conjecture that $N$ antibranes wrapping the ${\bf P}^1$
are also holographically dual to the conifold deformation with flux through the
$S^3$.  In fact, we have no choice but to require this as it is the result of acting by CPT
on both sides of the duality.  On the open string side, we replace the branes with
anti-branes, and on the closed string dual we have $N$ units of flux
through the $S^3$, but $N$ is now {\it negative}. In other words, the
superpotential is still given by \eff , but with $N < 0$.

At first sight, this implies that there is a critical point as given by
\crit, but now with $N$ negative.
However, this cannot be right, since $S$, which is the size of the
$S^3$, would grow without bound as we go to weak coupling $S\sim
\exp({ 8\pi^2 \over |N| {g_{\rm YM}}^2})$. This is clearly
unphysical, and moreover the description of the conifold breaks down
when $S$ is larger than the cutoff $\Lambda_0^3$ in the dual closed
string geometry.

To see what is going on, recall that on the open string
side the background has ${\cal N}=2$ supersymmetry, and adding D5
branes preserves an ${\cal N}=1$ subset of this, while adding
anti-branes preserves an orthogonal ${\cal N}=1$ subset.
By holography, we should have the same situation on the other side of
the duality.  Namely, before turning on flux, the background has
${\cal N}=2$ supersymetry, and turning on flux should break this
to ${\cal N}=1$. But now holography implies that depending on whether
we have branes or anti-branes in the dual -- so depending
on the sign of $N$ -- we should have one or the other ${\cal N}=1$ subgroup
of the supersymmetry preserved.
\lref\Antoniadis{
  I.~Antoniadis, H.~Partouche and T.~R.~Taylor,
``Spontaneous Breaking of N=2 Global Supersymmetry,''
  Phys.\ Lett.\ B {\bf 372}, 83 (1996)
  [arXiv:hep-th/9512006].}
It is clear that the superpotential \eff\ has been adapted to a superspace
in which the manifest ${\cal N}=1$ supersymmetry is the one preserved by branes,
and hence is {\it not} well adapted for the supersymmetry of the anti-branes.
Nevertheless, the theory with negative flux should somehow find a way to
capture the supersymmetry, as string holography predicts!  We will
now show that this is indeed the case.

The vacua of the theory are clearly classified by the critical
points of the physical {\it potential} $V$ of the theory:
$$
\del_S V=0,
$$
where
$$
V=g^{S\bar S}\,|\del_S{\cal W}|^2.
$$
Above, $g_{S\bar S}$ is the Kahler metric for $S$.  In this case, the
theory has softly broken ${\cal N}=2$ supersymmetry, and so $g$ is
given in terms of the prepotential as in the ${\cal N}=2$ case
$$
g_{S\bar S} = {\rm Im}(\tau),
$$
where
$$
\tau(S) = \del^2_S {\cal F}_0 = {1\over 2 \pi i} \,\log (S/\Lambda_0^3)
$$
With superpotential ${\cal W}$ as given in \supcon, the effective
potential becomes
$$
V = {2 i \over (\tau -\bar \tau)}\;|\alpha+ N \tau|^2
$$
It is easy to see that the critical points are at
$$
\del_S V = -{2 i\over (\tau-\bar\tau)^2}\, {\del_S^3} {\cal F}_0\,
({\bar \alpha}+N {\bar \tau})\,(\alpha+N{\bar \tau})=0.
$$
This has two solutions.  The first is at
\eqn\br{
\alpha + N \tau = 0
}
which solves $\del {\cal W} =0$,
and corresponds to \crit. It is physical when $N$ is positive.
The second critical point is at
\eqn\abr{
\alpha + N {\bar \tau} = 0.
}
Note that in this vacuum, $\del {\cal W} \neq 0$. In terms of $S$, it corresponds to
\eqn\crittwo{ \vev{S} = {\Lambda_0}^{3} \exp({{2 \pi i
\over |N|}{\bar \alpha}(\Lambda_0)})\; }
and is the physical vacuum when $N$ is {\it negative} (i.e. where we have $|N|$
anti-branes).

So how can it be that even though $\partial W \not= 0$,
supersymmetry is nevertheless preserved?  Toward understanding this,
recall that before turning on flux, the
closed string theory has ${\cal N}=2$
supersymmetry with one ${\cal N}=2$ $U(1)$ vector multiplet ${\cal A}$.
Adding D5 branes in the original theory corresponds to turning
on positive flux,
which forces this to decompose into
two ${\cal N}=1$ supermultiplets:
a chiral multiplet ${\cal S}$ containing $S$ and its superpartner $\psi$,
and a vector multiplet ${W_{\alpha}}$ containing the $U(1)$
gauge field
(coming from the 4-form potential
decomposed in terms of the harmonic 3-form on
$S^3$ and a 1-form in 4d) and the gaugino $\lambda$:
$$
{\cal A} = ({\cal S}, W^{\alpha})
$$
where
$$
\eqalign{
\cal{S} =  & \,S+\,\theta\,\psi+\ldots\cr
W_\alpha = &\, \lambda_\alpha+{i\over 2} (\sigma^{\mu\nu}\theta)_{\alpha} F_{\mu \nu}+\ldots.
}
$$
%
The Lagrangian in ${\cal N}=1$ superspace is given by
$$
{\cal L}={\cal L}_0+{\cal L}_{\cal W}
$$
where
$$
{\cal L}_0=Im\left(\int d^2\theta
d^2\bar\theta \; \overline{{\cal S}}_i{\del{\cal F}_0\over\del{\cal
S}_i}+\int d^2\theta{1\over2}{\del^2{\cal F}_0\over\del{\cal
S}_i\del{\cal S}_j}W_i^{\alpha}W_{j\alpha}\right)
$$
is the action for an ${\cal N}=2$ supersymmetric theory with
prepotential ${\cal F}_0$,
and the superpotential term is
$$ {\cal
L}_{\cal W}=\int d^2\theta{\cal W}({\cal S})+c.c.
$$
where the superpotential ${\cal W}({\cal S})$ is given by \supcon\ and is repeated here for the reader's convenience
$$
{\cal W}(S) = \alpha \; S +N\;
{\partial \over \partial S} {\cal F}_0.
$$
The puzzle is now what happens in the anti-D5 brane case.  Since the
Lagrangian with flux apparently has only ${\cal N}=1$
supersymmetry preserved by the D5 branes for $any$ value of $N$,
positive or negative, how is it possible that the anti-brane preserves
a different ${\cal N}=1$?  One might guess that despite the flux,
the Lagrangian ${\cal L}$ actually preserves the full ${\cal N}=2$
supersymmetry. This is too much to hope for. In particular,
turning on flux should be holographically dual to adding in the branes,
which $does$ break half of the ${\cal N}=2$ supersymmetry of the background.  Instead,
it turns out that the flux breaks ${\cal N}=2$ supersymmetry 
in a rather exotic way.  Namely, $which$ ${\cal N}=1$
is preserved off-shell turns out to be a choice of a ``gauge'': we can write the theory in a way which makes either the brane or the antibrane ${\cal N}=1$ 
supersymmetry manifest, no matter what $N$ is.  On
shell however we have no such freedom, and only one ${\cal N}=1$
supersymmetry can be preserved.  Which one this is depends only on whether
the flux is positive or negative, and not on the choice of the ${\cal N}=1$
supersymmetry made manifest by the lagrangian.  

To see how all this comes about, let us try to make
the spontaneously broken ${\cal N}=2$ supersymmetry manifest.
%
The ${\cal N}=2$ vector multiplet ${\cal A}$ is
really a chiral multiplet, satisfying the ${\cal N}=2$ chiral constraint ${\bar D}_{i\alpha} {\cal A}=0$,
\eqn\supper{
{\cal A} = S +  \theta^i \Psi_i + \theta^i \theta^j X_{ij}
+ {1\over 2} (\epsilon_{ij}{\theta}^i \sigma^{\mu \nu} \theta^j) F_{\mu\nu} +\ldots
}
Here $i,j=1,2$ are the $SU(2)_R$ indices, and $\Psi_i$ is a
doublet of fermions:
$$
\Psi=\pmatrix{\psi\cr\lambda}.
$$
The auxiliary fields $X_{ij}$ of each ${\cal N}=2$ chiral superfield form
an $SU(2)_R$ triplet,
satisfying a reality constraint
\eqn\cond{
{\bar X}^{ij}= \epsilon_{il} \epsilon_{jk} X^{lk}\equiv X_{ij}.
}
In particular,
$X^{11} = \bar{X}^{22} = X_{22}$, and so on.\foot{More precisely,
${\cal A}$ is a $reduced$ ${\cal N}=2$ chiral multiplet,
meaning it satisfies an additional constraint:
$
D^{ij}D_{ij}{\cal A} \sim
\nabla^2 \bar{\cal A},$
where
$D_{ij} = D_{i\alpha}D_{j}^{\alpha}$,
and $\nabla^2$ is the standard laplacian.
The reducing constraint says that
$
\nabla^2\epsilon_{il} \epsilon_{jk} X^{lk}= \nabla^2 {\bar X}^{ij}.
$
This implies that we can shift 
$X$ by a $constant$ imaginary part that does not satisfy \cond .
}

The action ${\cal L}$ can be written in terms of ${\cal N}=2$
superfields, where turning on fluxes in the geometry
corresponds to giving a vacuum expectation value
to some of the ${\cal N}=2$ F-terms \vafa\ref\APT{ I.~Antoniadis, H.~Partouche and T.R.~Taylor, ``Spontaneous
Breaking of N=2 Global Supersymmetry,'' CPTH-S397.1195 -- LPTENS-95/55
(2006) [arXiv:hep-th/9512006].  }\ref\IZ{E.A.~Ivanov and B.M.~Zupnik,
``Modified N=2 Supersymmetry and Fayet-Iliopoulos Terms,'' JINR
E2-97-322 (1997) [arXiv:hep-th/9710236]}.
Our presentation here follows closely \IZ . Namely, consider
the Lagrangian
\eqn\more{
{\rm Im}\left( \int d^2\theta_1d^2\theta_2\;\, {\cal F}_0({\cal A})\,\right)
+X_{ij} E^{ij} + X^{ij} E_{ij}.
}
where $E^{ij}$ is the triplet of Fayet-Iliopolous terms, with same properties as $X$ has.
Since the $X_{ij}$ transform by total derivatives, the FI term $X_{ij} E^{ij}$
preserves the ${\cal N}=2$ supersymmetry.
This will match ${\cal L}$ precisely if we set
$$
E^{11} = {\alpha} = {\bar E}^{22}, \qquad E^{12} = 0
$$
and moreover, we give
$X_{ij}$ a non-zero imaginary part
\eqn\shift{
X_{ij} \rightarrow X_{ij} + i N_{ij}.
}
where
\eqn\ft{
N_{11}\, =\,  0,\,\qquad   N_{22}=2\,N , \qquad N_{12} = 0.
}
It is easy to see from \supper\ that to decompose ${\cal A}$ in terms of  
${\cal N}=1$ multiplets, we can simply expand it in powers of $\theta_2$
\eqn\standard{
{\cal A}(\theta_1,\theta_2) = {\cal S}(\theta_1) +
\theta_2^{\alpha}\, W_\alpha(\theta_1) +\theta_2\theta_2
\;G(\theta_1)
}
where the chiral multiplet $G$ is given by (see, for example \ref\lykken{J.~D.~Lykken,
``Introduction to supersymmetry,''
arXiv:hep-th/9612114.
})\foot{Explicitly,
$$
\theta_2\theta_2 G(\theta_1) = X_{22} \theta_2 \theta_2 +
{1\over 2}(\epsilon_{ij}\theta^i\sigma_{\mu\nu}\theta^j)
(\theta^2 \sigma^{\mu \nu} \sigma^{\rho}\del_{\rho} \psi)
+{1\over 3!}(\epsilon_{ij}\theta^i\sigma_{\mu\nu}\theta^j)^2
 \nabla^2 {\bar S}.
$$
See for example \ref\deRooMM{
  M.~de Roo, J.~W.~van Holten, B.~de Wit and A.~Van Proeyen,
  ``Chiral Superfields In N=2 Supergravity,''
  Nucl.\ Phys.\ B {\bf 173}, 175 (1980).
}
}
$$
G(y,\theta_1)=
\int d^2
{\bar\theta}_1 \;{\bar {\cal S}}(y-i \theta_1\sigma{\bar \theta}_1, {\bar \theta}_1)= X_{22} + \ldots.
$$
Then, plugging in the vacuum expectation values above in \more\ and integrating over
$\theta^2,$ we recover the ${\cal N}=1$ form of ${\cal L}$. More precisely (this will get a nice interpretation later) we recover it, with the addition of a constant term $8\pi N/ g_{YM}^2$. Note that 
by shifting $X$ we have turned on a non-zero $F$
term in the $\theta^2$ direction {\it off-shell}. This breaks ${\cal N}=2$ supersymmetry,
leaving the Lagrangian with only ${\cal N}=1$ supersymmetry along $\theta^1$ direction.

Consider now the vacua of the
theory.\foot{Explicitly, in this language, the F-term potential becomes
$$
{1\over 2 i}\;\tau\; {\hat X}_{ij}\, {\hat X}^{ij}-
{1\over 2 i}\;{\bar \tau}\; {\hat {\bar X}}_{ij}\, {\hat {\bar X}}^{ij}
+ {\hat X}_{ij}\,E^{ij} + {\hat X}^{ij} \,E_{ij}.
$$
where indices are always raised and lowered with $\epsilon$ tensor, e.g.
${\hat X}^{ij} = \epsilon^{ik} \epsilon^{jl} {\hat X}_{kl}.$
Moreover, using reality properties of ${\hat X}$, it is easy to see that
${\hat {\bar X}}^{ij}= {\hat X}_{ij} - i (N_{ij} + N^{ij}),$ and the
result follows.}
Let us denote the full $F$-term as
$$
{\hat X}_{ij}=X_{ij} + i N_{ij}.
$$
Then it is easy to see that $X$ gets an expectation value,
%
%
%
\eqn\ftermone{
\vev{{\hat X}_{11}}
= {2 i \over \tau-\bar \tau}\,({\bar \alpha}+ N {\bar \tau})
}
and
\eqn\ftermtwo{
\vev{{\hat X}_{22}}
= {2 i \over \tau-\bar \tau}\,({\alpha}+ N {\bar \tau})
}
Now, recall that the $physical$ vacua depend on whether
the flux is positive or negative. In particular, in the brane vacuum,
where $N$ is positive, we have
$\alpha + N \tau =0$, so
$$
\vev{\hat X_{11}} = 0, \qquad \vev{\hat X_{22}} \neq 0.
$$
Now, since the supersymmetry variations of the fermions are
$$
\delta {\Psi}_i = i \,{\hat X}_{ij}\, \epsilon^j+\ldots
$$
It follows immediately that in the brane vacuum
$$
\delta \psi = 0, \qquad \delta \lambda \neq 0
$$
and $\lambda$ is the goldstino. The
unbroken supersymmetry pairs up $S$ and $\psi$ into a chiral field ${\cal S}$
and $\lambda$ and the gauge field into $W_{\alpha}$, as in \standard.

By contrast, in the anti-brane vacuum, $\alpha + N {\bar \tau} =0$, so
$$
\vev{\hat X_{11}}\neq 0, \qquad \vev{\hat X_{22}} = 0.
$$
Correspondingly, now $\psi$ is the goldstino:
$$
\delta \psi \neq 0, \qquad \delta \lambda = 0
$$
Now the unbroken supersymmetry
corresponds to pairing up $S$ and $\lambda$ into a chiral field
${\tilde{\cal S}}$, and ${\psi}$ is the partner of the gauge field in
${\tilde W}_{\alpha}$. In other words, now, it is natural to write
${\cal A}$ as
$$
{\cal A} = ({\tilde{\cal S}}, {\tilde W}_{\alpha})
$$
or, more explicitly:
\eqn\nonstandard{
{\cal A}(\theta_1,\theta_2) = {\tilde {\cal S}}(\theta_2) +
\theta_1^{\alpha}\, {\tilde W}_\alpha(\theta_2) +\theta_1\theta_1 \, {\tilde G}(\theta_2).
}

It should now be clear how it comes about
that even though
the Lagrangian ${\cal L}$ has only ${\cal N}=1$ supersymmetry,
depending on whether the flux $N$ is positive or negative,
we can have {\it different} ${\cal N}=1$ subgroups preserved on shell.
The supersymmetry of the Lagrangian was
broken since the flux shifted $X$ as in \shift . 
More precisely, supersymmetry was broken $only$
because this shift of $X$ could not be absorbed in the field redefinition
of $X$, consistent with \cond . Off shell, $X$ is allowed to fluctuate, but
because the shift is by $i N_{ij}$, which is not of the form  \cond ,
its fluctuations could not be absorbed completely and supersymmetry really is broken. However the shift of $X$ by $N_{ij}$ in \ft\ is indistinguishable from shifting $X$
by
\eqn\fttwo{
N_{11}\, =-2\,N,  \,\qquad   N_{22}=0 , \qquad N_{12} = 0.
}
since the difference between the two shifts $can$ be absorbed 
into a redefinition of the fields.  This would preserve a different ${\cal N}=1$ subgroup of the ${\cal N}=2$
supersymmetry, the one that is natural in the anti-brane theory.
Correspondingly, in all cases,
${\cal N}=2$ supersymmetry is broken to ${\cal N}=1$ already at the level of the Lagrangian. However, $which$ ${\cal N}=1$ is realized off-shell is a
gauge choice.

Next we compute the masses of bosonic and fermionic excitations around the vacua. The relevant terms in
the ${\cal N}=2$ Lagrangian are:
$$
 \int d^4 \theta\, {1\over 2}\, \del_{S}^3 {\cal F}_0
\;(\Psi_i \theta^i)\;(\Psi_j \theta^j)\;
({\hat X}_{kl} \theta^k\theta^{l})
$$
In the present context, we can simplify this as
$$
{1\over 2}\,\del_{S}^3 {\cal F}_0  (\Psi_1 \Psi_{1})\; {\hat X}_{22} +
{1\over 2}\,\del_{S}^3 {\cal F}_0 (\Psi_2 \Psi_{2})\; {\hat X}_{11}
$$
%
%
%
%
%
%
%
%
%
This gives fermion masses\foot{This is the physical mass, with canonically normalized kinetic terms.}
$$
m_{\psi} = {2i \over (\tau-\bar{\tau})^2}\; (\alpha+N {\bar
\tau}) \; \del_S^3 {\cal F}_0
$$
$$
m_{\lambda}= {2i \over (\tau - {\bar \tau})^2}\; ({\bar \alpha}
+ N {\bar \tau})\; \del_S^3 {\cal F}_0.
$$
For comparison, the mass
of the glueball field $S$ is\foot{ It is important to note here that
the kinetic terms for $S$ are fixed for us by the string large N
duality, and ${\cal N}=2$ supersymmetry.  They differ from the
``canonical'' kinetic terms of \VY\ . We are writing here the
physical masses, in the basis where all the fields have canonical
kinetic terms.}
$$
m_S = {2\over |\tau-\bar\tau|^2}\; \left( |\alpha+N {\tau}|^2+
|\alpha+N {\bar \tau}|^2\right)^{1\over 2}\; |\del_S^3 {\cal F}_0|.
$$
As a check, note that in the brane vacuum \br,
$\lambda$ is indeed massless as befits the
partner of the gauge field (since the original theory had ${\cal N}=2$ supersymmety, $\lambda$ here is in fact a goldstino!).
Moreover, it is easy to see that the masses of $\psi$ and $S$ agree,
as they should,
$$
|m_S| = {1\over 2\pi} {N^2 \over |S|\, |{\rm Im}(\alpha)|} = |m_\psi|, \qquad
m_{\lambda} =0.
$$
%

%

Now consider the case of
anti-branes where, in the holographic dual, we have negative flux.
This corresponds to the vacuum \abr .
The gauge field $A$ is still
massless and $S$ is massive.
{}From \abr , it follows that in this vacuum it is
$\psi$ which is massless, and $\lambda$ becomes massive.
Moreover, the mass of $\lambda$ is the same as for $S$.
$$
|m_S| ={1\over2\pi}{N^2 \over |S|\, |{\rm Im}(\alpha)|} = |m_\lambda|, \qquad
m_{\psi} =0.
$$
This is all beautifully consistent with holography!

To summarize, the anti-brane vacuum preserves different supersymmetry than
the brane vacuum, and the same is true in the large $N$ dual.
In other words the anti-brane/negative flux has oriented the
${\cal N}=1$
supersymmetry differently,
%
%
as is expected based on the holographic duality.

Let us now try to see if we can understand the anti-brane gluino condensate
\crittwo\ directly from the gauge theory.
Before adding in branes, the background has ${\cal
N}=2$ supersymmetry. The corresponding
supercharges are two weyl fermions $Q$,
${\tilde Q}$ that transform as a doublet under the $SU(2)_R$ symmetry
of the theory.  In either vacuum with the branes, only the $U(1)_R$
subgroup of the $SU(2)_R$ symmetry is preserved.
If adding in D-branes preserves $Q$, then anti-D
branes will preserve ${\tilde Q}$. Since $Q$ and ${\tilde Q}$
transform $oppositely$ under the $U(1)_R$ symmetry, in going from brane
to anti-brane, the chirality of the world-volume fermions flips.

The superpotential $\tilde {\cal W}$ that is an honest $F$ term with respect to the
supersymmetry preserved by the anti-brane theory, and which reproduces \crittwo, is
$$
{\tilde {\cal W}} =
{\overline{\alpha}}\; {S} + {1\over2\pi i}\, N\, S (\log( S/{\Lambda_0^3})-1),
$$
%
%
where $S$ is the anti-brane gaugino condensate, and $N$ is negative.
This reflects the fact that, if we keep the background fixed,
then $\theta$ transforms under the $R$ symmetry still as
\thetaeq , but the gauginos transform oppositely from
before. The rest is fixed by holography -- in other words,
$\alpha$ is the good chiral field which has $\theta$ as a component. Since it is
${\tilde{\cal W}}$ that is the F-term in this case, it is this, and not ${\cal W}$,
that should be independent of scale for the anti-brane, and hence ${\alpha}$ should
run as
$$
 2\pi i \alpha(\Lambda_0) = -\log ({\overline{\Lambda\over \Lambda_0}})^{-3N}
$$
Note that this is consistent with turning on the other F-term vev \fttwo ,
which directly preserves the anti-brane supersymmetry.

Finally, consider the value of the potential in the minimum.
At the brane vacuum, supersymmetry is unbroken, and
$$
V^{\rm brane}=0
$$
At the anti-brane vacuum, with negative $N$, we have:
$$
V^{\rm anti-brane} = -16 \pi {N\over g_{YM}^{2}}=16\pi{|N|\over g_{YM}^2}
$$
This treats branes and anti-branes asymmetrically. It is natural then to shift the
zero of the potential and have
$$V_* = V+8 \pi {N\over g_{YM}^{2}}$$
which leads to the minimum value
$$
V_*^{\rm anti-brane} =  8 \pi {|N|\over g_{YM}^{2}} = V_*^{\rm brane}
$$
This is equivalent to the inclusion of the constant term which came from the 
magnetic FI term in the ${\cal N}=2$ theory.  Note that this corresponds to the tension 
of $|N|$ (anti-)branes which is predicted by holography!  We thus find a beautiful confirmation
for the stringy anti-brane holography.

\newsec{Geometric metastability with branes and anti-brane}

There are many ways to naively break supersymmetry in string theory, but they
typically lead to instabilities. The simplest example of such a
situation is found in type IIB string compactifications on a
Calabi-Yau, where one introduces an equal number of $D3$ branes and
anti-branes filling the spacetime, but located at points in the
Calabi-Yau.  Supersymmetry is clearly broken, since branes and
anti-branes preserve orthogonal subsets of supersymmetry.
This does not typically lead to metastability, however, as one can move branes to where
the anti-branes are located, and such a motion would be induced by an attractive
force.  This is then similar to a familiar problem in classical
electrostatics, and is plagued by the same difficulty as one has in
finding locally stable equilibrium configurations of static electric charges.

One way to avoid such an obstacle, as was suggested in \vafa , is instead of
considering branes and anti-branes occupying points in the Calabi-Yau,
to consider wrapping them over minimal cycles inside the Calabi-Yau.
Moreover, we choose the minimal cycles to be isolated.  In other words
we assume there is no continuous deformations of the minimal cycles.
We wrap branes over a subset $C_1$ of such cycles and we wrap the
anti-branes over a distinct subset $C_2$.  In the case of compact Calabi-Yau, it is necessary that
the class $[C_1]+[C_2]=0$, so that there is no net charge (this condition can be modified if we have orientifold planes).
Because the branes and anti-branes cannot move without an increase in
energy, these will be a metastable equilibrium configurations (see Fig. 2).
This is true as long as the cycles $C_1$ and $C_2$ are separated by
more than a string scale distance, so that the geometric reasoning remains valid
(when they are closer there are tachyon modes which cause an
instability).

\bigskip
\centerline{\epsfxsize 4.0truein\epsfbox{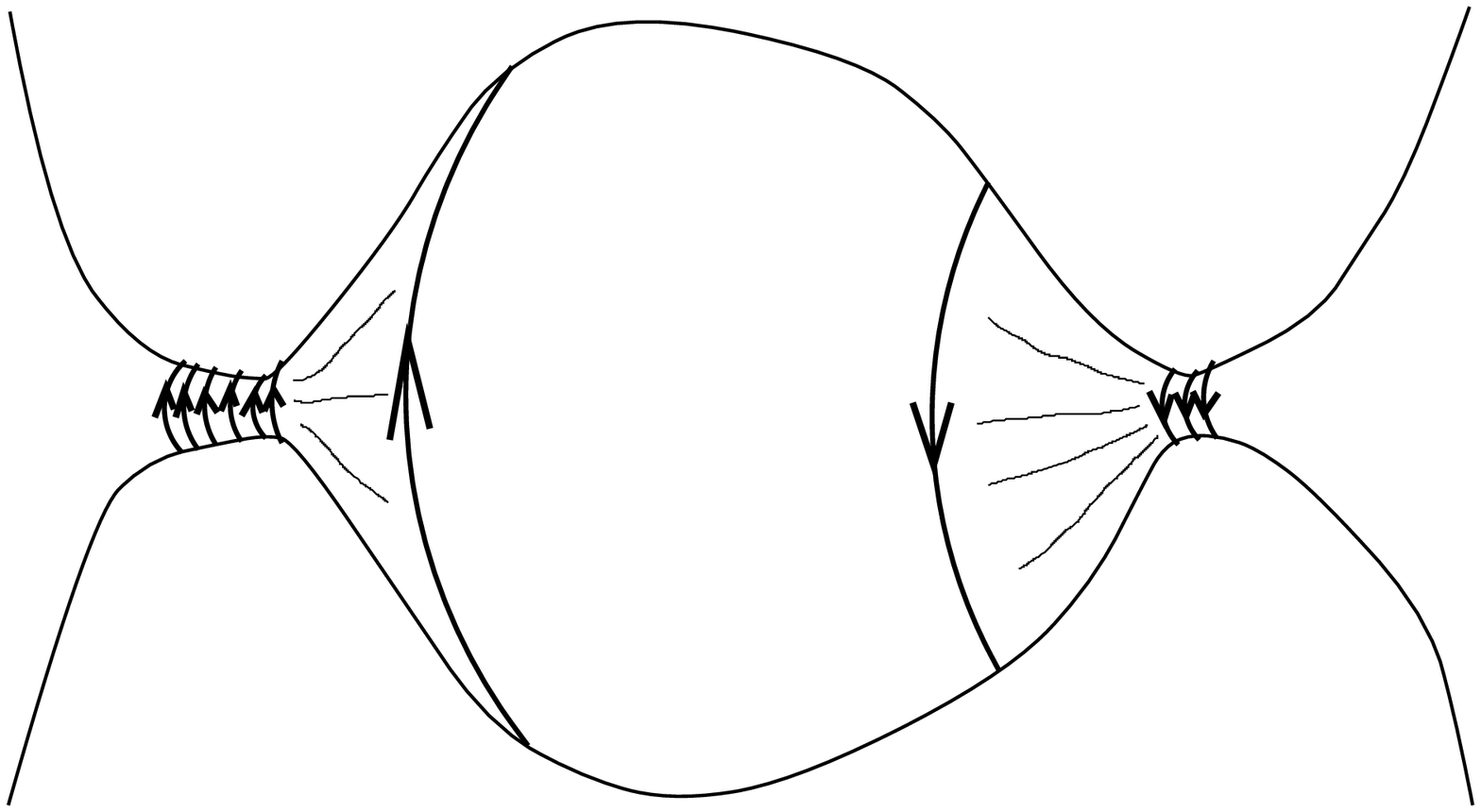}}
\noindent{\ninepoint
\baselineskip=2pt {\bf Fig. 2.} {Before antibranes and branes can annihilate one other, they will have to move and meet somewhere, and thus they will have to increase their volume due to the Calabi-Yau geometry.}}
\bigskip

Here we will consider non-compact examples of this
scenario. This decouples the lower dimensional gravity from the
discussion, and moreover, the condition that the net class
$[C_1]+[C_2]$ be zero is unnecessary as the flux can go to
infinity.  In particular, we will consider type IIB strings on a
non-compact Calabi-Yau threefold where we can wrap 4d space-time
filling D5 and anti-D5 branes over different isolated $S^2$'s in the
Calabi-Yau.  The local geometry of each $S^2$ will be the resolved
conifold reviewed in the previous section.  The only additional point
here is that we have more than one such $S^2$ in the same homology
class.  Moreover, we will consider the large $N$ limit of such
brane/anti-brane systems and find that the holographically dual closed
string geometry is the same one as in the supersymmetric case with
just branes, except that some of the fluxes will be negative.  This
leads, on the dual closed string side, to a metastable vacuum with
spontaneously broken supersymmetry.

We will first describe general geometries of this kind which support
meta-stable configurations with both D5  and anti-D5
branes. The relevant Calabi-Yau geometries turn out to be the same
ones studied in \civ , which
led to a non-perturbative formulation of the dual geometry in terms
of matrix models \dv.  The new twist
is that we now allow not just branes, but both branes and anti-branes to be present.
We will describe the holographically dual flux vacua.  We then specialize
to the case of just two $S^2$'s with branes wrapped over one $S^2$ and anti-branes wrapped
over the other, and study it in more detail.

\subsec{Local multi-critical geometries}

Consider a Calabi-Yau manifold given by
\eqn\cyb{ uv = y^2+W'(x)^2 }
where
$$
W'(x) = g \prod_{k=1}^n (x-a_k).
$$
If, for a moment, we set $g=0$, then this is an $A_1$ ALE singularity
at every point in the $x$ plane. One can resolve this by blowing up,
and this gives a family of ${\bf P}^1$, i.e. holomorphic 2-spheres,
parameterized by $x$.  Let
us denote the size of the blown up ${\bf P}^1$ by $r$. In string theory, this
is complexified by the $B^{NS}$-field\foot{On the world volume of the D5 brane
$B_{NS}$ gets mixed up with the RR potential, so more precisely by
$r/g_s$ we will mean the complex combination of the kahler modulus with
$B^{RR}+\tau B^{NS}$. To not have the dilaton turned on, we will in fact
keep only the later, setting the geometric size to zero \klebs .}. Turning $g$
back on lifts most of the singularities, leaving just $n$ isolated ${\bf P}^1$'s at
$x=a_k$.  Of course, the $S^2$ still exists over each point $x$, but
it is not represented by a holomorphic ${\bf P}^1$. In other words, its
area is not minimal in its homology class.  We
have
\eqn\area{A(x)=(|r|^2+|W'|^2)^{1/2}.}
where $A(x)$ denotes the area of the minimal $S^2$ as a function of
$x$.  This can be seen by the fact that the equation for each $x$ is
an ALE space and the above denotes the area of the smallest $S^2$
in the ALE geometry for a fixed $x$.  Even though the minimal
${\bf P}^1$'s are now isolated, i.e.  at points where $W'=0$, they are
all in the same homology class, the one inherited from the ALE
space\foot{ As explained in \GVW , the parameters $g$ and $a_i$
which enter \cyb\ and $W$ are $not$ dynamical fields, but enter in
specifying the theory.  This is possible because the Calabi-Yau is non
compact.}.  Note also that the first term, $|r|^2$, does not depend on
$x$.

One can now consider wrapping some number of branes $N_k$, $k=1,\ldots
n$ on each ${\bf P}^1$.
The case when all $N_k$'s are positive and all
the branes are D5 branes was studied in \civ.
In that case, the gauge theory on the branes is 
an ${\cal N}=1$ supersymmetric $U(N)$ gauge theory with
an adjoint matter field $\Phi$ and superpotential
given by
\eqn\suppotree{{\rm Tr} W(\Phi)}
where $W$ is the same polynomial whose derivative is given by
$W'(x)$ in the defining equation of Calabi-Yau  \civ \ref\KKe{
    S.~Kachru, S.~Katz, A.~E.~Lawrence and J.~McGreevy,
  ``Open string instantons and superpotentials,''
  Phys.\ Rev.\ D {\bf 62}, 026001 (2000)
  [arXiv:hep-th/9912151]
  M.~Aganagic and C.~Vafa,
  ``Mirror symmetry, D-branes and counting holomorphic discs,''
  arXiv:hep-th/0012041.
  F.~Cachazo, S.~Katz and C.~Vafa,
  ``Geometric transitions and N = 1 quiver theories,''
  arXiv:hep-th/0108120.
}.
     The eigenvalues of $\Phi$ are identified
with positions of the D5 brane on the $x$-plane.
Here
$$N=\sum_k N_k$$
and the choice of distribution of D5 branes among the critical
points $a_k$ corresponds to the choice of a Higgs branch of
this supersymmetric gauge theory where, the gauge group
is broken
$$U(N)\rightarrow \prod_{k=1}^n U(N_k)$$
In the low energy limit we have ${\cal N}=1$ Yang-Mills
theories $U(N_k)$ with $\Phi$ field corresponding to a massive
adjoint matter for each.
Note that \suppotree\ is consistent
with \area\ in the large $r$ limit,  because wrapping a brane over ${\bf P}^1$ and considering
its energy as a function of $x$, it is minimized along with the area at points where $W'(x)=0$.
Furthermore, it is clear from this that the effective coupling constant $g_{\rm YM}$ of the four dimensional gauge theory living
on the brane is the area of the minimal ${\bf P}^1$'s times  $1/g_s$:
$${1\over g_{\rm YM}^2}={|r|\over g_s}$$
%
\bigskip
\centerline{\epsfxsize 3.0truein\epsfbox{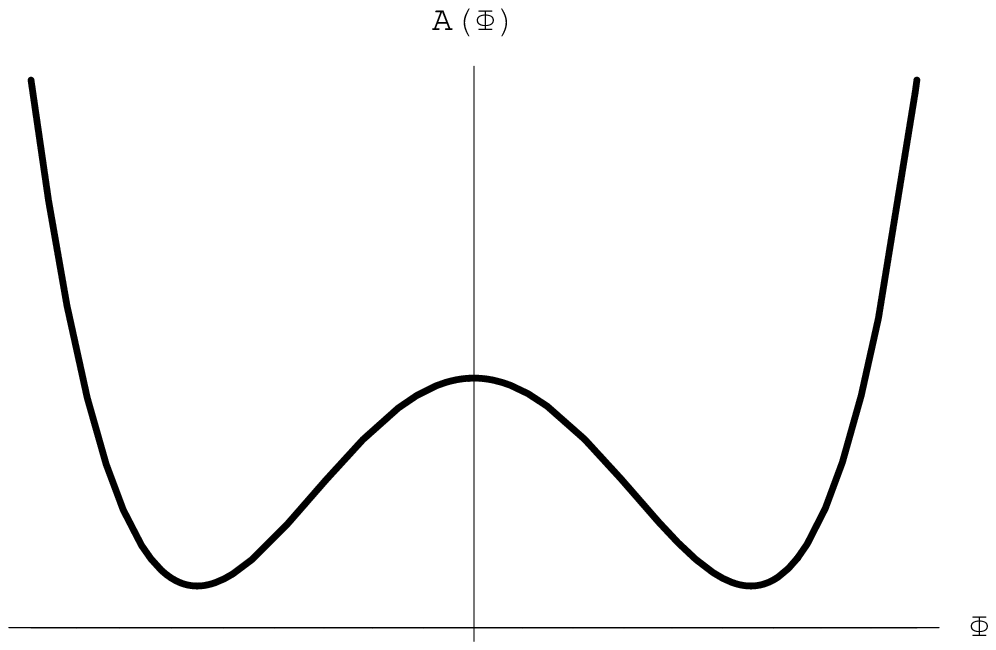}}
\noindent{\ninepoint
\baselineskip=2pt {\bf Fig. 3.} { The brane theory with a tree level
physical potential $g_s V(x)= {|r|} + {1\over 2 r } |dW|^2 
\approx A(x)$ is depicted
along a real line in the $x$-plane for the case where $W$ has two critical points. The potential
reflects the Calabi-Yau geometry, and
the leading term
represents the brane tension.}} 
\bigskip

 Below, we will study
what happens when some of the ${\bf P}^1$'s are wrapped with D5 branes and
others with anti-D5 branes.  
By making the vacua $a_i$ very widely
separated, the branes and the anti-branes should interact very
weakly. Therefore, we still expect to have an approximate
${\cal N}=1$ supersymmetric gauge theory for
each brane, with gaugino condensation and
confinement
at low energies, just as discussed in the previous section.  However,
because
supersymmetry is broken and there are lower energy vacua available
where some of the branes annihilate, the system should be only
meta-stable.  Note that the fact that the ${\bf P}^1$'s are isolated
but in the same class is what guarantees metastability. For the branes
to annihilate the anti-branes, they have to climb the potential
depicted in Fig. 3.  This should be accurate
description of the potential when the branes are far away from each other,
where the effect of supersymmetry breaking is small.  When the branes
and anti-branes are very close together -- for example when
they are within a string distance -- there would be tachyon
in the theory and the above potential will not be an accurate
description.  Nevertheless it is clear that the minimum of a brane and anti-brane
system is realized when they annihilate each other.
We have
thus geometrically engineered a metastable brane configuration which breaks supersymmetry.
We will discuss aspects of the open string gauge dynamics of
this configuration in section 3.6. As we
will discuss in that section, unlike the supersymmetric case, there seems
to be no simple field theory with a finite number of degrees of freedom
which captures the brane/anti-brane geometry.  Of course, one
can always discuss it in the context of open string field theory.

We have a control parameter for this metastability,
which also controls the amount of supersymmetry breaking, which
is related
to the separation of the critical points $a_i$.  The farther apart 
they are, the more stable our system is.  We will discuss 
stability and decay rate issues in section 4.  In the next
subsection, we study the large $N$ holographic dual for this system.

\subsec{The large $N$ dual description}

The supersymmetric configuration of branes for this geometry was
studied in \civ, where a large $N$ holographic dual was proposed.  The
relevant Calabi-Yau geometry was obtained by a geometric transition of
\cyb\ whereby the ${\bf P}^1$'s are blown down and the $n$
resulting conifold singlularities at $x=a_k$ are resolved into $S^3$'s by 
deformations of the complex structure. It is given by
\eqn\cya{ uv = y^2+W'(x)^2 + f_{n-1}(x), }
where $f_{n-1}(x)$ is a degree $n-1$ polynomial in $x$.  As
explained in \civ, the geometry is effectively described by a
Riemann surface which is a double cover of the $x$ plane, where the
two sheets come together along $n$ cuts near $x=a_k$ (where the
${\bf P}^1$'s used to be).  The $A$ and $B$ cycles of the
Calabi-Yau project to 1-cycles on the Riemann surface, as in
Fig. 4. The geometry is characterized by the periods of the
$(3,0)$ form $\Omega$,
$$
S_k= \oint_{A^k} \Omega ,\qquad \del_{S_k} {\cal F}_0=\int_{B_k} \Omega.
$$
%
\bigskip
\centerline{\epsfxsize 6truein\epsfbox{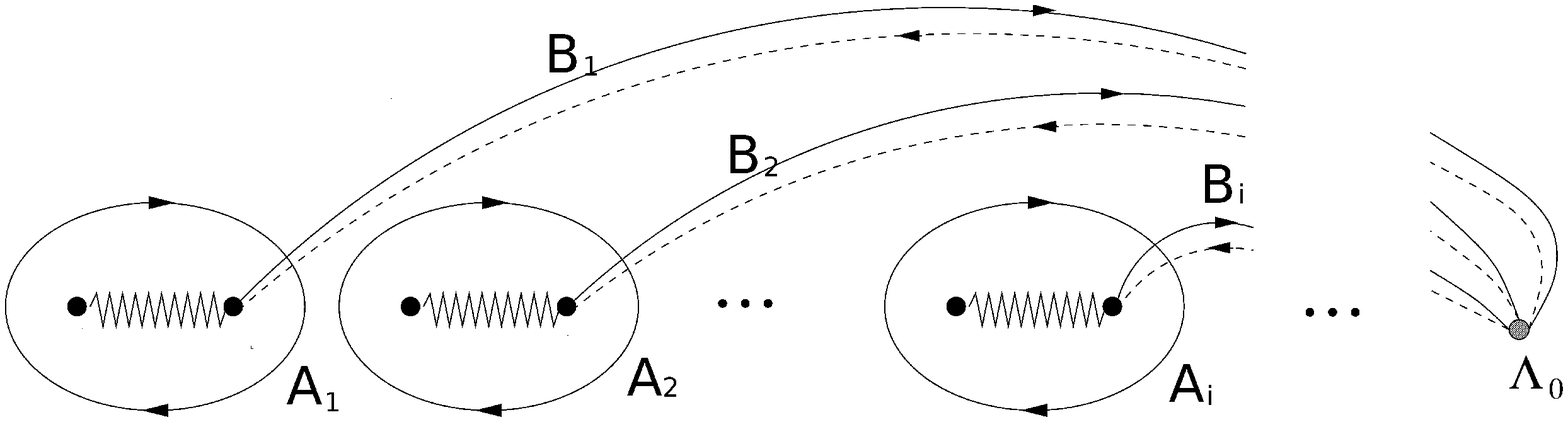}}
\noindent{\ninepoint \baselineskip=2pt {\bf Fig. 4.}{ The multi-critical
Calabi-Yau geometry projected to the $x$-plane.  The cut structure reflects the branching
of $y$ over the $x$ plane for $u=v=0$.} }
\bigskip
The complex scalars $S_k$ are the scalar partners of the $n$ $U(1)$ vector
multiplets under the ${\cal N}=2$ supersymmetry, and ${\cal F}_0(S)$ is the
corresponding prepotential. As before, while \cya\ depends also on $a_i$
and $g$, the latter are just fixed parameters.

If, before the transition, all of the ${\bf P^1}$'s were wrapped with a large number
of branes, the holographically dual type IIB string theory is given by
 the \cya\ geometry, where the branes from before the transition
are replaced by fluxes
$$
\oint_{A^k} H = N_k, \qquad \int_{B_k} H = -\alpha.
$$
We want to conjecture that this duality holds whether or not
all the $N_k\geq 0$.  In fact, the discussion of the previous section
shows that if all the $N_k \leq 0$ the duality should continue to hold
by CPT conjugation.  Our conjecture is that it also holds as
when some $N_k$ are positive and some negative.  The flux numbers
$N_k$ will be positive or negative depending on whether we had
D5 branes or anti-D5 branes wrapping the $k$'th ${\bf P^1}$ before the
transition.  The flux through the $B_k$ cycles will correspond to the
bare gauge coupling constant on the D-branes wrapping the
corresponding ${\bf P}^1$, and is the same for all $k$ as the ${\bf
P}^1$'s are all in the same homology class.  Turning on fluxes generates a
superpotential
\GVW\
$$
{\cal W} =\int H\wedge \Omega,
$$
\eqn\superpotential{{\cal W}(S) = \sum_k \alpha
S_k + N_k {\del_{S_k}} {\cal F}_0. }

In \civ , in the case when all $N_k$'s are positive, the ${\cal N}=1$ chiral superfield corresponding to
$S_k$ is identified with the gaugino condensates of the $SU(N_k)$
subgroup of the $U(N_k)$ gauge group factor,
$S_k =  {1\over 32 \pi^2} {\rm Tr}_{SU(N_k)} W_{\alpha} W^{\alpha}$,
before the transition. When we have both branes and anti-branes, as
long as they are very far separated, this picture should persist.
Namely, the brane theory should still have gaugino condensation and
confinement, and $S_k$'s should still correspond to the gaugino
condensates, even though we expect the supersymmetry to be completely
broken in this metastable vacuum. Moreover, for each $k$, there is
a remaining massless $U(1)$ gauge field in the dual geometry. It gets
identified with the
massless $U(1)$ on the gauge theory side which is left over by the
gaugino condensation that confines the $SU(N_k)$ subgroup of the
$U(N_k)$ gauge theory.

In the supersymmetric case studied in \civ, the coefficients
of the polynomial $f(x)$ determining the dual geometry and the sizes of $S_i$ is fixed by the
requirement that
$$\partial_{S_k}W(S) =0$$
and this gives a supersymmetric holographic dual.  In the case of interest
for us, with mixed fluxes, we do not expect to preserve supersymmetry.
Instead we should consider the physical potential $V(S)$ and find the dual geometry
by extremizing
$$\partial_{S_k}V(S)=0$$
which we expect to lead to a metastable vacuum.
The effective potential $V$ is given in terms of the
special geometry data and the flux quanta:
$$
V = g^{S_i{\bar S}_j}\,
{\del_{S_i}}{\cal W}\,\overline{\del_{S_j}{\cal W}}
$$
where the Kahler metric is given by
$
g_{i {\bar j}} = {\rm{Im}}(\tau_{ij})
$
in terms of the period matrix of the Calabi-Yau
$$
\tau_{ij} = {\del_{S_i} \del_{S_j}} {\cal F}_0.
$$
In terms of $\tau$, we can write
\eqn\thpot{
V = \left({2 i \over \tau -\bar{\tau}}\right)^{jk} (\alpha_j +
\tau_{jj'} N^{j'}) ({\bar \alpha}_k + {\bar \tau}_{kk'} N^{k'})
}
where we have all the $\alpha_j=\alpha$.
To be explicit, we consider in detail the case where we have only two $S^3$'s
in the dual geometry before turning to the more general case.

\subsec{More precise statment of the conjecture}

Let us make our conjecture precise: We conjecture that the large $N$
limit of brane/anti-brane systems are Calabi-Yaus with fluxes.  In
particular, the solutions to tree level string equations on the closed
string side lead to an all order summation of planar diagrams of the
dual brane/anti-brane system (i.e. to all orders in the `t Hooft
coupling).  We translate this statement to mean that the geometry of
the closed string vacuum at tree level is captured by extremizing
the physical potential.  Moreover the physical potential is
characterized by the fluxes (which fix the superpotential) as well as
by the Kahler potential; the main ingredient for both of these objects
is the special geometry of the Calabi-Yau after transition.  This, in turn,
is completely fixed by tree level topological string theory, or equivalently by
the planar limit of a certain large $N$ matrix model \dv .  Here we are using
the fact that since ${\cal N}=2$ is softly broken by the flux
terms, the special Kahler metric is unaffected at the string tree
level.  It is quite gratifying to see that topological objects, such
as matrix integrals, play a role in determining the geometry of
non-supersymmetric string vacua!

Of course the Kahler potential should be modified at higher string
loops since.  In particular the $1/N$ corrections to our duality should
involve such corrections.  Note that, in the supersymmetric case
studied in \civ, there were no $1/N$ corrections to modify the
geometry of the vacua.  We do expect the situation to be different in
the non-supersymmetric case.  Note however, 
from the discussion of section 2, the soft breaking
is such that it is ambiguous $which$ ${\cal N}=1$ supersymmetry 
the Lagrangian has. This should constrain what kind of quantum corrections one can have beyond those allowed by a generic soft breaking. This
deserves further study.

\subsec{The case of $2$ $S^3$'s}

For simplicity we start with the case where we have just two $S^3$'s.
Before the transition, there are two shrinking ${\bf P^1}$'s at
$x=a_{1,2}$.  Let us denote by $\Delta$ the distance between them,
$$
\Delta = a_1-a_2.
$$
The theory has different vacua
depending on the number of branes we put on each ${\bf P}^1$.
The vacua with different
brane/antibrane distributions are separated by energy
barriers. To overcome these, the branes must first become more massive.

The effective superpotential of the dual geometry, coming from the electric and magnetic FI terms turned on by the fluxes, is
$$
{\cal W}(S) = \alpha (S_1+ S_2) + N_1 {\del_{S_1}}
{\cal F}_0+N_2 {\del_{S_2}} {\cal F}_0
$$
%
%
%

%
The B-periods have been computed explicitly in \civ .
We have
\eqn\pero{ \eqalign{ 2 \pi i \del_{S_1}{\cal F}_0=
W(\Lambda_0)-W(a_1) + S_1 (\log({S_1\over g\Delta^3})-1)-
&2(S_1+S_2)\log({\Lambda_0\over \Delta}) \cr
+&(2 S_1^2-10S_1 S_2 + 5 S_2^2)/(g \Delta^3) +\ldots}
}
and
\eqn\pert{ \eqalign{ 2 \pi i \del_{S_2}{\cal F}_0=
W(\Lambda_0)-W(a_2) + S_2 (\log({S_2\over g \Delta^3})-1)-
&2(S_1+S_2)\log({\Lambda_0\over \Delta}) \cr -&(5 S_1^2-10S_1 S_2 +
2 S_2^2)/(g \Delta^3) +\ldots} }
where the omitted terms are of order $S_1^{n_2} S_2^{n_2}/(g\Delta^3)^{n_1+n_2-1}$,
for $n_1+n_2>2$. In the above,
$\Lambda_0$ is the cutoff used in computing the periods of the non-compact $B$ cycles,
and physically corresponds to a high energy cutoff in the
theory.

To the leading order (we will justify this aposteriori),
we can drop the quadatic terms in the ${S_k \over g \Delta^3}$'s, and higher.
To this order,
$$
\eqalign{
2\pi i \tau_{11}
=&
2 \pi i\, {\del^2_{S_1}} {\cal F}_0
\approx \log({S_1 \over g\Delta^3})- \log({\Lambda_0 \over \Delta})^2 \cr
2\pi i \tau_{12}
= &
2 \pi i\, {\del_{S_1}\del_{S_2}} {\cal F}_0
\approx -\log({\Lambda_0\over \Delta})^2
\cr
2\pi i \tau_{22}
= &
2 \pi i\, {\del^2_{S_2}} {\cal F}_0
\approx \log({S_2 \over g \Delta^3})- \log({\Lambda_0 \over \Delta})^2
}
$$
In particular, note that at the leading order ${\tau}_{12}$
is independent of the $S_i$, so we can use $\tau_{ii}$ as variables.
It follows easily that the minima of the potential are at
$$
Re(\alpha) + Re(\tau)_{ij} N^j =0
$$
and
$$
{\rm Im}(\alpha) + {\rm Im}(\tau)_{ij} |N^j| =0
$$
For example, with branes on the first ${\bf P}^1$ and anti-branes on the second,
$N_1>0>N_2$,
\eqn\nonsusy{
\vev{{S}_1} = g \Delta^3\; ({\Lambda_0\over \Delta})^{2}
(\overline{\Lambda_0\over \Delta})^{2|{N_2 \over N_1}|} \;e^{-2\pi i
\alpha/|N_1|}, \qquad \vev{{S}_2} = {{ g \Delta^3}} ({\Lambda_0\over
\Delta})^{2} (\overline{\Lambda_0 \over \Delta})^{2|{N_1 \over
N_2}|} e^{2\pi i {\overline \alpha}/|N_2|}.
}
To see how to interpret this, let us recall the
supersymmetric situation when
$N_1,N_2>0$. There, to the same order, one finds
\eqn\susy{
\vev{{S}_1} = g \Delta^3\;
({\Lambda_0\over \Delta})^{2(1+|{N_2\over N_1}|)}
\;e^{-2\pi i \alpha/|N_1|},
\qquad \vev{{S}_2} = {{ g \Delta^3}}
({\Lambda_0\over \Delta})^{2(1+|{N_1\over N_2}|)}
e^{-2\pi i { \alpha}/|N_2|}.
}
The interpretation of the above structure in the
supersymmetric case is as follows.  In the IR,
the theory flows to a product of two supersymmetric gauge theories with gauge group 
$U(N_1)\times U(N_2)$, where each $U(N_i)$ factor is characterized
by a scale $\Lambda_i$,
\eqn\vevs{
\vev{S_1}=\Lambda_1^3 ,
\qquad \vev{S_2}=\Lambda_2^3 .
}
Let us denote by $\alpha_{1,2}$ the
the bare coupling constants of the low energy theory $U(N_1)$ and
$U(N_2)$ theories, i.e.
\eqn\defcou{
2 \pi i \alpha_{i} = - \log(\,{\Lambda_i\over \Lambda_0}\,)^{3 |N_i|},
\qquad i=1,2
}
On the one hand, we can simply read them off from \susy\
since we can write \vevs\ as
$\vev{S_i} = \Lambda_0^3 e^{-2\pi i \alpha_i/ |N_i|}$:
\eqn\sue{\eqalign{
2\pi i \alpha_{1} &= 2\pi i \alpha +
\log({\Lambda_0\over m_\Phi})^{|N_1|}-2\log({\Lambda_0\over m_W})^{|N_2|}\cr
2\pi i \alpha_{2} &= 2\pi i \alpha +
\log({\Lambda_0\over m_\Phi})^{|N_2|}-2\log({\Lambda_0\over m_W})^{|N_1|}.
}
}
We could also have obtained this by using the matching relations with
the high energy coupling $\alpha$.  Namely, in flowing from high
energies we have to take into account the massive fields we are
integrating out.  The fields being integrated out in this case are the
massive adjoint $\Phi$ near each vacuum, whose mass goes as $m_\Phi =g
\Delta$, and the massive $W$ bosons in the bifundamental representation,
whose masses are $m_W=\Delta$. The standard contribution of these heavy fields
to the running coupling constant is simply what is written in \sue .

In the non-supersymmetric case,
from \vevs\ and \defcou , we can write the gaugino vevs in \nonsusy\
very suggestively exactly as in \sue\ except that we replace
$$
\Lambda_0^{|N_1|+|N_2|}\rightarrow
\Lambda_0^{|N_1|} {\overline
\Lambda_0}^{|N_2|}, \Lambda_0^{|N_2|} {\overline
\Lambda_0}^{|N_1|},
$$
and
$$
m_{W}\rightarrow {\overline m_{W}}
$$
leading to
\eqn\nsue{\eqalign{
2\pi i \alpha_{1} &= 2\pi i \alpha +
\log({\Lambda_0\over m_\Phi})^{|N_1|}-2
\log({{\overline \Lambda_0}\over {\overline m_W}})^{|N_2|} \cr
2\pi i \alpha_{2} &= {\overline{2\pi i \alpha}} +
\log({\Lambda_0\over m_\Phi})^{|N_2|}-2\log({{\overline \Lambda_0}\over {\overline m_W}})^{|N_1|}
}}
which gives back \nonsusy.
In other words, the leading result is as in the supersymmetric
case, except that branes of the opposite type lead to complex conjugate running.
In particular, this implies that the real part of the coupling
runs as before, but the $\theta$-angle is running differently
due to the matter field coming between open strings stretched between 
the branes and anti-branes.  We will discuss potential explanations of this in section 3.6,
where we discuss the matter structure in the brane/anti-brane
system.

The potential at the critical point is given by\foot{We have shifted the
overall
zero point energy of $V$
as we did in the previous section,
to make the brane and the anti-brane tensions equal. More precisely, we set
$$V\rightarrow V + {8 \pi \over g_{YM}^2}\, (N_1+N_2).$$
}
\eqn\pamin{
V_{*}^{+-} = {8 \pi \over g_{YM}^2}\, (|N_1|+|N_2|) -
{{2\over\pi} | N_1||N_2|}\; \log|{\Lambda_0 \over m_W}|^2
}
The first term, in the holographic dual, corresponds to the tensions
of the branes. The second term should correspond to the Coleman-Weinberg
one loop potential which is generated by zero point energies of the
fields.  This interpretation coincides nicely with the fact that this term is
proportional to $|N_1||N_2|$, and thus comes entirely from the $1-2$
sector of open strings with one end on the branes and the other on the anti-branes.
The fields in the $1-1$ and $2-2$ sectors with both open string endpoints on the same type of brane
do not contribute to this, as those sectors are supersymmetric and the boson and fermion
contributions cancel. We shall return to this in section 3.6.
For comparison, in the case of where both ${\bf P}^1$'s were wrapped by D5 branes,
the potential at the critical point $V_*^{++}$
equals
$$
V_{*}^{++} =
{8 \pi \over g_{YM}^2}\, (|N_1|+|N_2|)=V_{*}^{--}
$$
and is the same as for all anti-branes.  This comes as no surprise, since the tensions are the same,
and the interaction terms cancel since the theory is now truly
supersymmetric.

We now consider the masses of bosons and fermions in the
brane/anti-brane background. 
With supersymmetry broken, there
is no reason to expect the $4$ real bosons of the theory, coming from
the fluctuations of $S_{1,2}$ around the vacuum, to be pairwise degenerate.  
To compute them, we simply expand the
potential $V$ to quadratic order. More precisely, to compute the
physical masses as opposed to the naive hessian of the potential, we have to
go to the basis where kinetic terms of the fields are canonical.  The computation
is straightforward, if somewhat messy. At the end of the day, we get
the following expressions.
$$
\left( m_\pm (c) \right)^2 = {(a^2 +b^2 +2abcv)\pm \sqrt{(a^2 +b^2
+2abcv)^2 -4a^2b^2(1-v)^2} \over 2 (1-v)^2} $$
where $c$ takes values $c=\pm 1$,
and
$$a \equiv \left|{N_1 \over 2\pi \Lambda_1^3 {\rm Im} \tau_{11}}\right|, \qquad
b \equiv \left|{N_2 \over 2\pi  \Lambda_2^3 {\rm Im} \tau_{22}}\right|$$
$$v \equiv {({\rm Im} \tau_{12})^2 \over {\rm Im} \tau_{11} {\rm Im}
\tau_{22}}.
$$
Indeed we find that our vacuum is metastable, because
all the $m^2>0$ (which follows
from the above formula and the fact that $v<1$ in the
regime of interest $|S_i/g\Delta^3|<1$).  This is a nice 
check on our holography
conjecture, as the brane/anti-brane construction was clearly metastable.  Moreover, we see that there are
four real bosons, whose masses are generically non-degenerate, as expected for the spectrum
with broken supersymmetry.

Since supersymmetry is completely broken from
${\cal N}=2$ to ${\cal N}=0$, we expect to find $2$ massless Weyl fermions,
which are the Goldstinos. More precisely, we expect this $only$
at the closed string tree
level (i.e at the leading order in $1/N$ expansion).
Namely, at string tree level, we can
think about turning on the fluxes as simply giving them an
``expectation value'', which would break the ${\cal N}=2$
supersymmetry spontaneously.
However, at higher orders, the Kahler potential should a priori
$not$ be protected
in that breaking is soft, and not spontaneous. This will affect the computation
of the mass terms in section 2, which relied on the ${\cal N}=2$
relation betweeen the Kahler potential and the superpotential, and should result in only one massless fermion remaining.
Masses of the fermions are computed from the ${\cal N}=2$
Lagrangian, as in section two, but now with two vector
multiplets.  We again redefine the fields so as to give them
canonical kinetic terms.
We indeed find two massless fermions, at the order at which we are working 
at. Since supersymmetry is broken
these are interpreted as the Goldstinos (we will give a more general argument in the next subsection).
There are also two massive fermions, with masses
$$
m_{\rm f1} = {a\over 1-v}, \qquad
m_{\rm f2} = {b\over 1-v}, \qquad
$$
Note that $v$ controls the strength of supersymmetry breaking.
In particular when $v\rightarrow 0$ the 4 boson masses become pairwise 
degenerate and agree with the two fermion masses $a$ and $b$, 
as expected for a 
pair of ${\cal N}=1$ chiral multiplets.

The mass splitting between bosons and fermions is a measure
of the supersymmetry breaking. In order for supersymmetry breaking to be 
weak, these splittings have to be small. 
There are two natural ways to make supersymmetry
breaking be small:  One way  is to 
take the number
of anti-branes to be much smaller than the number of
branes.  The other way is to make the branes and anti-branes
be very far from each other.  We will consider mass splittings
in both of these cases.

Adding a small anti-brane charge should be a small
perturbation to a system of a large number of branes.  In that
context, the parameter that measures supersymmetry breaking should be
$|N_2/N_1|$, where $|N_2|$ is the number of anti-branes.
Let us see how this is reflected in $v$.  We should see that in this
limit, $v$ becomes small.

Note that \nonsusy\ implies that

$$
\left|{ {S_1} \over g \Delta^3} \right|^{|N_1|} = \left|{
{S_2} \over g \Delta^3} \right|^{|N_2|}
= \left|{ {\Lambda}
\over \Delta} \right|^{2(|N_1|+|N_2|)} =\exp\left(-(1-\delta){8
\pi^2 \over g_{\rm YM}^2}\right)
$$

with
$$\delta \equiv \left({g_{\rm YM} \over 2\pi}\right)^2
{(|N_1|+|N_2|)}\log \left|{ {\Lambda_0} \over \Delta} \right|.$$
Note that
\eqn\usefu{\left|{\delta \over 1-\delta}\right|=\left| {\log \Lambda_0/\Delta \over \log 
\Delta/\Lambda}\right| }

For finite non-vanishing $S_i/g\Delta^3$, clearly $\delta$ is very close to $1$ because of the fact 
that $g_{YM}^2$ is very small for large cutoff $\Lambda_0$.  Of course, this is
compatible with the definition of $\delta$ as we recall
the running of $g_{YM}^2$.
Therefore, let us write
$$\delta= 1-\epsilon$$
We then have (without any approximation)

\eqn\usev{v={({\rm Im} \tau_{12})^2 \over {\rm Im} \tau_{11} {\rm Im}
\tau_{22}}
={(1-\epsilon)^2\over (1+\epsilon^2) +
\epsilon (\left|{ N_2\over 
N_1}\right|+ \left| {N_1\over N_2}\right|)}}
And in the limit of ${|N_1| \over |N_2|} \rightarrow \infty$ (or
similarly when ${|N_2| 
\over 
|N_1|} \rightarrow \infty$) we find that $v \rightarrow 0$.
This is as expected, since in this limit we expect
supersymmetry breaking to be very small.

Below are plots of the masses for nonsupersymmetric vacua, with fixed $|N_1
N_2|$ but varying $\log \left| {N_2 \over N_1} \right| $. They
demonstrate how mass splitting vanishes in the
extreme limits of the ratio $N_1/N_2$, where supersymmetry
is restored.

\bigskip
\centerline{\epsfxsize 6truein\epsfbox{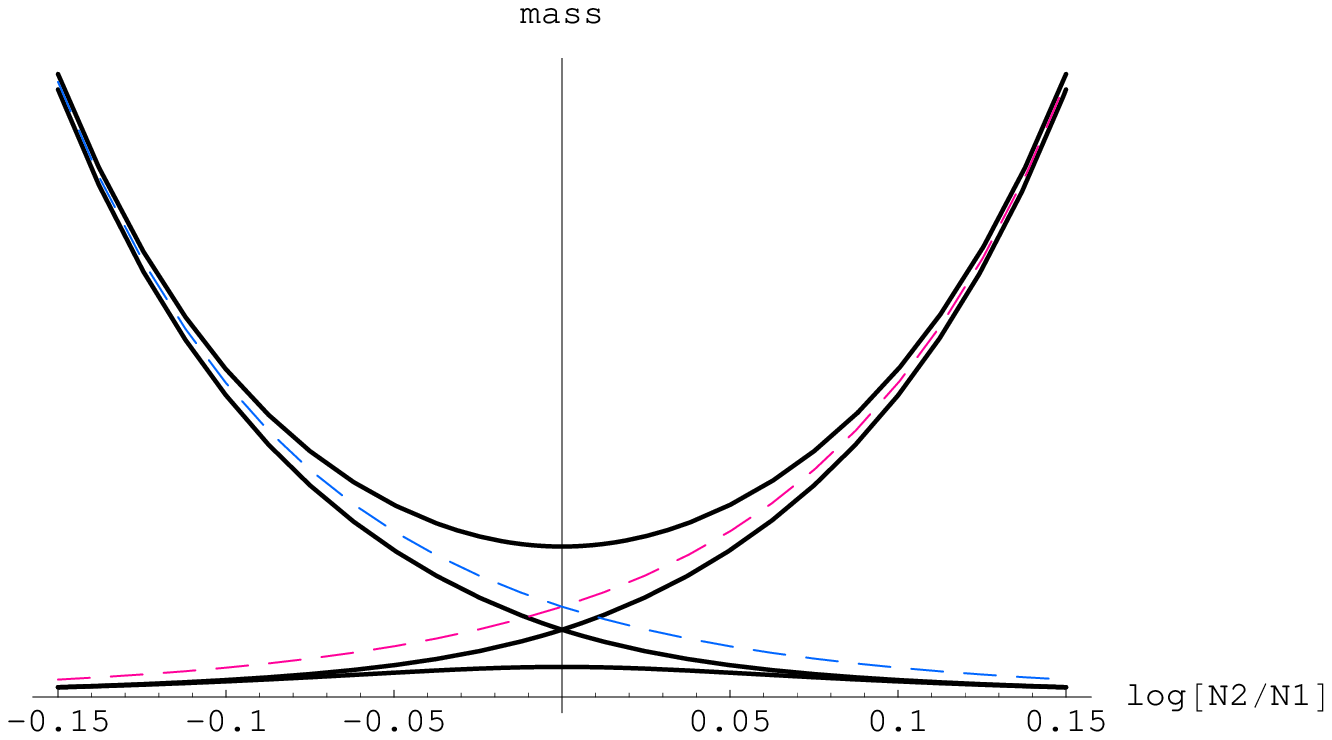}}
\noindent{\ninepoint \baselineskip=2pt {\bf Fig. 5.} {Here we depict
the masses of the four bosons as solid lines and the two
non-vanishing fermionic masses as dashed lines. The other two
fermions, not shown here, are massless Goldstinos. Note that
supersymmetry breaking is most pronounced when $|N_1|=|N_2|$. This
plot is for fixed $|N_1 N_2|$, as we vary $|N_1/N_2|$.}}
\bigskip

It is also natural to consider the mass splittings in the
limit of large separation of branes, where supersymmetry is
expected to be restored.  We have to be somewhat careful
in taking this limit.  It turns out that the right limit is
when
$$|\Lambda| <<|\Delta| <<|\Lambda_0|$$
and where
$$|\Delta/\Lambda |>> |\Lambda_0/\Delta|$$
which implies from \usefu\ that $\delta\rightarrow 0$ and from \usev\ we have
$v\rightarrow 0$ and thus the mass splittings disappear as expected.

It is also interesting to write the energy of the vacuum \pamin\ as a
function of $v$.  In particular let $\Delta E$ represent
the shift in energy from when the branes are infinitely far away
from one another.  Then we find from \pamin\ that
$${\Delta E\over E}=-  {\delta\over (\sqrt{|N_1| \over 
|N_2|}
+\sqrt{|N_2| \over |N_1|})^2}$$
Note that the energy shift goes to zero as $|N_2/N_1|\rightarrow 0,\infty$ as
expected.  It also goes to zero in the large brane/anti-brane separation,
because then $\delta \rightarrow 0$.
The sign of the shift is also correct, as
one would expect that the attraction between the branes 
and the anti-branes should would decrease the energy of the system.
Note also that for fixed $|N_1N_2|$, $\Delta E/E$, as well as the mass splittings, is maximal
when $N_1=-N_2$, i.e. when we have equal number of branes and anti-branes,
whereas if one of them is much larger than the other, mass splittings as well as $\Delta E/E\rightarrow 0$.
This is again consistent with the fact that in the limit with an
extreme population imbalance of branes, the supersymmetry breaking goes away.

\subsec{$n$ $S^3$'s}

Let us now generalize our analysis to the case with an arbitrary number of blown up $S^3$'s.  In this case, before the geometric transition there are $n$ shrinking ${\bf P}^1$'s located at $x=a_i$,  $i=1,2,\ldots,k$.  We will denote the distance between them as
$$
\Delta_{ij}=a_i-a_j
$$
As in the simpler case, for a specified total number of branes, we can classify the vacua of the theory by the distribution of the branes among the ${\bf P}^1$'s, and these vacua will be separated by potential barriers corresponding to the increase in the size of the wrapped ${\bf P}^1$ between critical points.  

The effective superpotential of the dual geometry after the geometric transition, in which the branes have been replaced by fluxes, is given by \cya.  The B-periods will now take the form
$$
2\pi i\del_{S_i}{\cal F}_0=S_i\log{S_i\over W''(a_i)\Lambda_0^2}+\sum_{j\neq i}S_j\log\left({\Delta_{ij}^2\over\Lambda_0^2}\right)+\ldots
$$
where again, $\Lambda_0$ is interpreted as a UV cutoff for the theory, and we are omitting terms analogous to the polynomial terms in the example of the last section.

To leading order, the generalization of the 2 $S^3$ case to the current situation is straightforward.  We have
$$
\eqalign{
2\pi i \tau_{ii}
=&
2 \pi i\, {\del^2_{S_i}} {\cal F}_0
\approx \log\left({S_1 \over W^{\prime\prime}(a_i)\Lambda_0^2}\right)\cr
2\pi i \tau_{ij}
= &
2 \pi i\, {\del_{S_i}\del_{S_j}} {\cal F}_0
\approx -\log\left({\Lambda_0^2\over \Delta_{ij}^2}\right)
}
$$
The potential (3.7) has its critical points where
\eqn\critn{\eqalign{
\del_{S_i} V =& {i\over2}{\cal F}_{ijl} g^{S_jS_j'} g^{S_kS_k'}
(\bar{\alpha}_j +{\bar\tau}_{jj'}N^{j'}) (\alpha_k
+{\bar\tau}_{kk'}N^{k'})=0
}}
The solutions which correspond to minima are then determined by
$$
Re(\alpha) + Re(\tau)_{ij} N^j =0
$$
and
$$
{\rm Im}(\alpha) + {\rm Im}(\tau)_{ij} |N^j| =0
$$
Indeed, in the case where all $N^i$ are either positive (negative), these are just the (non-)standard F-flatness conditions of the ${\cal N}=1$ supersymmetric theory, so they will be satisfied exactly.  In the non-supersymmetric cases, the conditions will receive perturbative
corrections coming from the corrections to the Kahler potential at higher
string loops.

The expectation values of the bosons $S^i$ take a natural form in these vacua, being given by
$$
\eqalign{
\vev{S_{i}}=&W^{\prime \prime }(a_{i})\Lambda _{0}^{2} \prod_{j\not=i}^{N_j
>0}\left( { \Lambda _{0}  \over  \Delta _{ij}}\right) ^{2{|N_{j}|
\over |N_{i}|}}\prod_{k\not=i}^{N_k <0} {\left( \overline{  \Lambda
_{0} \over \Delta _{ik}}\right) }^{2{|N_{k}| \over |N_{i}|}} \exp
\left( -{2\pi i\alpha  \over |N_{i}|}\right),\qquad N_i>0\cr
\vev{S_{i}}=&W^{\prime \prime }(a_{i})\Lambda _{0}^{2} \prod_{j\not=i}^{N_j
>0}\left( \overline{ \Lambda _{0}  \over  \Delta _{ij}}\right) ^{2{|N_{j}|
\over |N_{i}|}}\prod_{k\not=i}^{N_k <0} {\left( \Lambda
_{0} \over \Delta _{ik}\right) }^{2{|N_{k}| \over |N_{i}|}} \exp
\left( {2\pi i\overline{\alpha}  \over |N_{i}|}\right),\qquad N_i<0\
}
$$
The explanation of these expectation values is the natural
generalization of the case from the previous section.  We characterize
each gauge theory in the IR by a scale $\Lambda_i$,
where $S_i=\Lambda_i^3$.  The massive adjoints near the $i$'th
critical point now have $m_{\Phi_i}=W^{\prime\prime}(a_i)$, whereas
the massive $W$ bosons have mass ${m_{W_{ij} }=\Delta_{ij}}$.  Then
matching scales, we find
$$
\eqalign{
\vev{S_{i}^{{\rm brane}}}^{\left|N_i\right|}=\Lambda_i^{3\left|N_i\right|}=&e^{-2\pi i\alpha}\Lambda_0^{2N_{brane}}\overline\Lambda_0^{2N_{\rm antibrane}}m_{\Phi_i}^{|N_i|} \prod_{j\neq i}^{\rm same}{m_{W_{ij}}}^{-2N_j}\prod_{j\neq i}^{\rm opposite}\overline {m_{W_{ij}}}^{-2N_j}\cr
\vev{S_{i}^{{\rm antibrane}}}^{\left|N_i\right|}=\Lambda_i^{3\left|N_i\right|}=&e^{2\pi i\overline\alpha}\overline\Lambda_0^{2N_{\rm brane}}\Lambda_0^{2N_{\rm antibrane}}m_{\Phi_i}^{|N_i|} \prod_{j\neq i}^{\rm same}{m_{W_{ij}}}^{-2N_i}\prod_{j\neq i}^{\rm opposite}\overline {m_{W_{ij}}}^{-2N_j}\cr
}
$$
Above, $N_{\rm brane}$ and $N_{\rm anti-brane}$ are the total number of
branes and anti-branes, respectively, and the products are over the
sets of branes of the same and opposite types as the brane in
question. In the supersymmetric case, this reduces to the
expected relation between the scales.  The non-supersymmetric case is,
to leading order, identical except that the branes of opposite type
contribute complex conjugate running, which is explained in section
3.6.

Similarly to the 2 $S^3$ case, the gauge coupling constant in each
$U(N_k)$ factor will run as
$$
2\pi i \alpha_{i} = 2\pi i \alpha +
\log({\Lambda_0\over m_{\Phi_i}})^{|N_i|}-2  \sum_{j\neq i}^{\rm same}\log({\Lambda_0\over m_{W_{ij}}})^{|N_j|}-2\sum_{j\neq i}^{\rm opposite}
\log({{\overline \Lambda_0}\over {\overline m_{W_{ij}}}})^{|N_2|},
$$
for a brane, ($N_i >0$).
For antibrane, ($N_i <0$) and the running is
$$
2\pi i \alpha_{i} = \overline{2\pi i \alpha} +
\log({\Lambda_0\over m_{\Phi_i}})^{|N_i|}-2  \sum_{j\neq i}^{\rm same}\log({\Lambda_0\over m_{W_{ij}}})^{|N_j|}-2\sum_{j\neq i}^{\rm opposite}
\log({{\overline \Lambda_0}\over {\overline m_{W_{ij}}}})^{|N_2|}.
$$
This again has a simple form corresponding to integrating out
massive adjoints and $W$ bosons. Moreover, the fields corresponding to strings with both endpoints on branes, or on the anti-branes, contribute to running of the gauge coupling as in the supersymmetric case.
The fields coming from strings with one end on the brane and the other
on the anti-brane, again give remarkably simple contributions, almost as in the supersymmetric case: the only apparent difference being in
how they couple to the $\theta$ angle. We will come back to this later.

One could in principle compute
the masses of the bosons and fermions in a generic one of these vacua.
We will not repeat this here. Instead, let us just try to
understand the $massless$ fermions, or the structure of supersymmetry breaking in this theory.
The results of section 2 are easy to generalize, to find the fermion
mass matrices\foot{Note that the mass matrices are now being expressed
not in the basis in which the kinetic terms take the canonical form,
but in the basis in which the Lagrangian was originally expressed.  This will not affect our analysis.}
\vskip 0.1cm
$$
m_{\psi^i\psi^j} =- {i \over2} g^{S_l {\bar S}_{k}} {\cal
F}_{ijl} (\alpha_k +{\bar\tau}_{kk'}N^{k'})
$$
\vskip 0.1cm
$$
m_{\lambda^i\lambda^j} = -{i \over2} g^{S_l {\bar S}_{k}} {\cal
F}_{ijl}({\bar \alpha}_k +{\bar\tau}_{kk'}N^{k'}).
$$
In the above, the repeated indices are summed over, and we have denoted
$${\cal F}_{ijk}={\partial^3 {\cal F}_0 \over \partial S_i \partial S_j \partial S_k}.$$
At the order to which we are working (corresponding to 
one loop on the field theory side), ${\cal F}_{ijk}$ is diagonal.
The masses then imply that in vacua where the standard F-term
$$
\del_{S_i} {\cal W} = \alpha_i + \tau_{ij} N^j
$$
fails to vanish, a zero eigenvalue is generated for the
$\psi$ mass matrix, while when the non-standard F-term
$$
\del_{S_i} {\tilde {\cal W}} = {\overline \alpha_i} + {\tau_{ij}} N^j
$$
fails to vanish, a zero eigenvalue is generated
for the $\lambda$ mass matrix.

More generally, note that the $exact$ conditions (at leading order in $1/N$) for critical points of the potential
{\critn} can be re-written in terms of the fermion mass matrices as
\eqn\mpsi{
m_{\psi^i\psi^j} g^{S_jS_k}({\bar \alpha}_k +
{\bar \tau}_{kk'}N^{k'})=0
}
\eqn\mlambda{
m_{\lambda^i\lambda^j} g^{S_jS_k}({\alpha}_k +
{\bar \tau}_{kk'}N^{k'})=0.
}
In the case when supersymmetry is unbroken, and all the ${\bf
P^1}$'s are wrapped by branes, we will find both of these equations
to be truly exact, and satisfied trivially: the $\lambda$ mass
matrix vanish identically, so \mlambda\ is trivial, and \mpsi\ is
satisfied without constraining $m_{\psi^i\psi^j}$ directly since the
vanishing of $({\alpha}_k + {\tau}_{kk'}N^{k'})$ is the F-flatness
condition. A similar story holds in the anti-brane vacuum.

In the case with both branes and antibranes, the equations are $not$
trivial. Instead, they say that both the $\psi$- and the
$\lambda$-mass matrices have at least one zero eigenvalue. 
At the string tree
level (i.e at the leading order in $1/N$ expansion), we expect them to
have $exactly$ one zero eigenvalue each. To this order, we can
think about turning on the fluxes as simply giving them an ``expectation value'', which would break the ${\cal N}=2$ supersymmetry spontaneously.
However, at higher orders, the Kahler potential is $not$ protected
in that the symmetry breaking is soft, but not spontaneous.  Consequently, the
form of the equations \mpsi,\mlambda\ should presumably receive corrections,
and only one massles fermion should remain,
corresponding to breaking ${\cal N}=1$ to ${\cal N}=0$.

The leading-order potential in these vacua is given by
$$
V_*={8\pi\over {g_{\rm YM}}^2}\left(\sum_i\left|N_i\right|\right)-\left(\sum_{i,j}^{N_i>0,N_j<0}{2\over\pi}\left|N_i\right|\left|N_j\right|\log\left|{\Lambda_0\over\Delta_{ij}}\right|\right)
$$
This takes the natural form of a brane-tension contribution for each
brane in the system, plus interaction terms between the
brane/anti-brane pairs.  If we define
$$\delta_{ij} \equiv \left({g_{\rm YM}\over 2\pi}\right)^2
{(\sum_k |N_k|)}\log \left|{ {\Lambda_0} \over \Delta_{ij}} \right|.
$$
Then in terms of these parameters $\delta_{ij}$, the physical potential takes the simpler form of
$$
V_*= {8\pi\over {g_{\rm YM}}^2} \left(   \sum_i\left|N_i\right|   - {\sum_{i,j}^{N_i>0,N_j<0} \delta_{ij} |N_i||N_j| \over  \sum_i\left|N_i\right|} \right) $$ so that the energy shift is

$${\Delta E \over E} =
{   -   \sum_{i,j}^{N_i>0,N_j<0}  \delta_{ij} |N_i||N_j|  \over (  \sum_i\left|N_i\right|  
)^2  }$$
Once again, the forces among the (anti-)branes cancel. Moreover, the branes and the anti-branes attract, which lowers the energy of the vacuum,
below the supersymmetric one. This energy shift will effectively vanish only when there is an extreme imbalance between number of the 
branes and antibranes, just as in the 2 $S^3$ case.

\subsec{Brane/anti-Brane gauge system}

It is natural to ask whether we can find a simple description
of the non-supersymmetric theory corresponding to the
brane/anti-brane system.  This certainly exists in the full
open string field theory.  However, one may wonder whether one
can construct a field theory version of this which maintains only
a {\it finite} number of degrees of freedom.  In fact,
it is not clear if this should be possible, as we will now explain.

Consider a simpler situation where we have a stack of $N$ parallel
D3 branes.  This gives an ${\cal N}=4$ supersymmetric $U(N)$ gauge
theory.  Now consider instead $(N+k)$ D3 branes and $k$ anti-D3 branes
separated by a distance $a$.  This is clearly unstable, and
will decay back to a system with $N$ D3 branes.  Can one
describe this in a field theory setup?  There have been attempts
\ref\attemp{
J.~A.~Minahan and B.~Zwiebach,
  ``Gauge fields and fermions in tachyon effective field theories,''
  JHEP {\bf 0102}, 034 (2001)
  [arXiv:hep-th/0011226];

  J.~A.~Minahan and B.~Zwiebach,
  ``Effective tachyon dynamics in superstring theory,''
  JHEP {\bf 0103}, 038 (2001)
  [arXiv:hep-th/0009246];

  J.~A.~Minahan and B.~Zwiebach,
  ``Field theory models for tachyon and gauge field string dynamics,''
  JHEP {\bf 0009}, 029 (2000)
  [arXiv:hep-th/0008231].  }\
along these lines (in connection with Sen's conjecture \ref\sen{
  A.~Sen,
  ``Tachyon condensation on the brane antibrane system,''
  JHEP {\bf 9808}, 012 (1998)
  [arXiv:hep-th/9805170].
}),
however no complete finite truncation seems to exist.  This system
is similar to ours.  In particular, consider the case with 2 $S^3$'s, with a total of $N$ D5 branes.  We know that we can have
various supersymmetric vacua of this theory in which
$$U(N)\rightarrow U(N_1)\times U(N-N_1)$$
where we have $N_1$ branes wrap one ${\bf P}^1$ and $N-N_1$
wrap the other.  However we also know that the theory with
$N$ D5 branes can come from a meta-stable vacuum with
$(N+k)$ D5 branes wrap around one ${\bf P}^1$ and $k$ anti-D5 branes
around the other.  This theory would have a $U(N+k)\times U(k)$
gauge symmetry. Clearly we need more degrees of freedom than are present in the
$U(N)$ theory in order to to describe this theory.  In fact, since $k$ can
be arbtirarily large, we cannot have a single system with
a finite number of degrees of freedom describing all such metastable
critical points.  Of course, one could imagine that there could
be a theory with a finite number of fields which decribes all such configurations up to a given
maximum $k$, but this does not seem very natural.  It is thus
reasonable to expect that string field theory would be needed to
fully describe this system.

Nevertheless we have seen, to leading order, a very simple
running of the two gauge groups in the non-supersymmetric
case, and it would be natural to ask if we can explain
this from the dual open string theory.  In this dual theory, where 
we have some branes wrapping the first
${\bf P}^1$ and some anti-branes the other, we have three
different types of sectors:  the 1-1, 2-2 and the 1-2 open
string sectors.  The 1-1 and the 2-2 subsectors are supersymmetric
and give rise to the description of the ${\cal N}=1$ supersymmetric
$U(N_i)$ theories for $i=1,2$, coupled to the adjoint matter fields $\Phi$
which are massive.  This part can be inherited from the supersymmetric
case, and in particular explains the fact that the running of
the coupling constant of the $U(N_i)$ from the massive field $\Phi$
is the same as in the supersymmetric case.  The difference
between our case and the supersymmetric case comes from the 1-2
subsector.  In the NS sector we have the usual tachyon mode
whose mass squared is shifted to a positive value, as long as
$\Delta$ is sufficiently larger than the string scale, due
to the stretching of the open string between the far separated ${\bf P}^1$'s.
We also have the usual oscillator modes.
In the Ramond sector ground state the only difference between
the supersymmetric case and our case is that the fermions, which
in the supersymmetric case is the gaugino partner of the massive $W_{12}$
boson, has the opposite chirality.  The fact that they have
the opposite chirality explains the fact that the $\theta$ term
of the gauge group runs with an opposite sign compared to the
supersymmetric case.  In explaining the fact that the norm
runs the same way as the supersymmetric case, the contributions
of the fermions is clear, but it is not obvious why the NS sector
contribution should have led to the same kind of running.  It would
be interesting to explain this directly from the open string theory
annulus computation in the NS sector.

It is easy to generalize the discussion above to the $n$ $S^3$ case.
In fact just as in the 2 $S^3$ case, the only subsectors which are non-supersymmetric
are the $i-j$ sectors where $i$ is from a brane and $j$ is from an anti-brane cut.
This agrees with the results obtained in the general case, where supersymmetry
breaking contributions come, to leading order, precisely from these sectors.

\newsec{Decay Rates}

In the previous sections, we constructed non-supersymmetric vacua of
string theory corresponding to wrapping D5 branes and anti-D5 branes
on the 2-cycles of the a local Calabi-Yau.  Here, branes and anti-branes are
wrapping rigid 2-cycles which are in the same homology class, but are
widely separated, and there are potential barriers between them of tunable
height.  As we emphasized, the crucial aspect of this is
that the vacua obtained in this way can be long lived, and thus are
different from brane-anti brane systems in flat space, or systems of
branes and anti-branes probing the Calabi-Yau manifold, which have
been considered in the literature.  Moreover, when the
numbers of branes on each 2-cycle is large, this has a holographic dual in terms
of non-supersymmetric flux vacua. By duality, we also expect the
corresponding flux vacua to be metastable, despite breaking
supersymmetry.
In this section we will explore the stability of these vacua in more
detail.

In both open and closed string language, the theory starts out in a
non-supersymmetric vacuum. Since there are lower energy states
avaliable, the non-supersymmetric vacuum is a false vacuum.  As long
as the theory is weakly coupled throughout the process, as is the case
here, the decay can be understood in the semi-classical approximation.
This does not depend on the details of the theory, and we will
review some aspects of the beautiful analysis in \coleman .
Afterwards, we apply this to the case at hand.

The false vacuum decays by nucleating a bubble of true vacuum
by an instanton process.
The rate of decay $\Gamma$ is given in terms of the action of the relevant instanton as
$$
\Gamma \sim \exp(-S_I).
$$
The instanton action $S_I$ is the action of the euclidean bounce
solution, which interpolates between the true vacuum inside the bubble
and the false vacuum outside of the
bubble\foot{The coefficient of proportionality comes from the one loop
amplitude in the instanton background, and as long as $S_I$
is large, its actual value does not matter.}.
Assuming that the dominant instanton is a spherical bubble of radius $R$,
the euclidian instanton action is given by \coleman\
$$
S_I = -{\pi^2 \over 2} R^4 \Delta V + {2 \pi^2 R^3} S_D
$$
The first term is the contribution to the action from inside the
bubble.  Here $\Delta V$ is the difference in the energy density
between the true and the false vacuum, and this is multiplied by the
volume of the bubble (a 4-sphere of radius $R$). The second term is the
contribution to the action from the domain wall that interpolates
between the inside and the outside on the bubble, assuming the domain
wall is thin. There, $S_D$ is the tension of the domain wall, and
$2\pi^2 R^3$ is its surface area (the area of a 3-sphere of radius
$R$).  The radius $R$ of the bubble is determined by energetics: The
bubble can form when the gain in energy compensates for the mass of
the domain wall that is created. The energy cost to create a bubble of
radius $R$ is
$$
E = -{4\pi \over 3} R^3 \Delta V + {4 \pi R^2} S_D
$$
where the first term comes from lowering the energy of the vacuum
inside the bubble, and the last term is the energy cost due to surface
tension of the domain wall. We can create a bubble at no energy cost
of radius
$$
R_* = 3 {S_D \over \Delta V}.
$$
Bubbles created with radia smaller than this recollapse. The bubbles created at $R=R_*$ expand indefinitely, as this is energetically favored. At any rate, the relavant instanton action is
$S_I(R_*)$, or
$$
S_I = {27\pi  \over 2}  {S_D^4 \over (\Delta V)^3}.
$$
Correspondingly, the vacuum is longer lived the higher the tension
of the domain wall needed to create it, and the lower the
energy splitting between the true and the false vacuum.

With this in hand, let us consider the decays of the vacua we found.
When the number of branes wrapping each ${\bf P}^1$ is small, the open
string picture is appropriate. Consider, for simplicity, the case
where the superpotential has only two critical points
$$
W'(x) = g (x-a_1)(x-a_2),
$$
and we have only a single brane at $x=a_1$ and a single
anti-brane at $x=a_2$. The effective potential that
either of these branes sees separately is
given simply by the size of the $2$-sphere it wraps
\eqn\mass{
A(x) = (|W'(x)|^2+ |r|^2)^{1/2}
}
where $|r|$ is the tension of the brane wrapping the minimal
${\bf P^1}$ (see section 3). The branes are, of course, also charged,
so there is an electric field flux tube (of the six-form potential)
running between the brane and the anti-brane,
along a 3-cycle corresponding to an $S^2$ being swept from $x=a_1$ to
$x=a_2.$  Despite the flux tube, the system is (meta)stable
if the potential barrier from \mass\ is high enough.
The decay process corresponds,
for example, to having the D5-brane tunnel under the energy
barrier from the vacuum at $x=a_1$ to the vacuum $x=a_2$, to
annihilate with the anti-D5-brane there.  More precisely,
what happens is that a bubble of true vacuum is created in $R^{3,1}$,
inside of which the branes are anihilated. The boundary of the bubble
is a domain wall which, in the thin wall approximation,
corresponds to a D5 brane wrapping the $S^2$ boundary of the bubble
in $R^3$ and the 3-cycle in the Calabi-Yau where the flux tube was.
The tension of the domain wall is the the size of the 3-cycle that the D5 brane wraps in the Calabi-Yau
times the tension of the 5-brane
$$
S_D ={1\over g_s} \int_{a_1}^{a_2} A(x) dx \approx
{1\over g_s}\left| W(a_2)-W(a_1) \right| ={1\over 3 g_s}\, \left| g \Delta^3 \right|
$$
where the second term in \mass\ gives vanishingly small contribution\foot{For us, the mass of the brane comes
from the B-field only, and the brane is heavy only because the string
coupling is weak.}
in the limit $\Delta\gg r$.  The difference in the cosmological
constants of the two vacua corresponds, to the leading order, to the
difference in tensions between the original configuration, with
$
V_{i}= 2 {|r| /g_s}
$
%
and the final one where the branes have annihilated
$
V_f = 0,
$
so the gain in energy is simply
$$
\Delta V = 2 { |r| \over g_s} = {2\over g_{YM}^2}
$$ 

When the number of branes wrapping the ${\bf P}^1$'s is large,
we use the dual geometry from after the transition, with fluxes.
Consider, for example, the case when $N_1$ branes wraps the first
${\bf P}^1$ and $|N_2|$ anti-branes the second, with $|N_{1,2}|$
large, and satisfying
$$
N_1>0>N_2.
$$
Then, the $C$ cycle that runs between the two cuts,
$[C] = [B_1] - [B_2]$ is a compact flux tube, with $|N_2|$ units of
$H^{RR}$ flux running through it.
To decay, one kills off the flux lines one by one, by creating an
D5 brane wrapping a zero size ${\bf P}^1$ where the first
${\bf P^1}$ used to be before the transition. This brane
grows and eats up the flux line until it vanishes again at the
first cut. After all the flux has decayed, this leaves us with
$N_1+N_2$ units of flux through the first $S^3$.
The tension of the corresponding domain wall is
$$
S_D = {|N_2|\over g_s} \int_{C} \left| \Omega \right|
$$
where
$$
\oint_C \Omega = \int_{B_1} \Omega - \int_{B_2} \Omega
\approx W(a_2) - W(a_1).
$$
%
%
where we omit terms that are exponentially suppressed in the
vacuum.\foot{More precisely, we should evaluate the corrections to this
at the values of $S_{1,2}$ correpsonding to being
somewhere in the middle between the true and the false vacuum.
Either way, the omitted terms are suppressed by, at
least, ${\Lambda_{1,2} \over \Delta}$, and we can neglect them.}
In this case, the energy of
the false vacuum is, to the leading order,
$
V_i =
(|N_1| + |N_2|){|r|/ g_s}.
$
After the flux has decayed, we have
$
V_f = (|N_1|- |N_2|){|r|/g_s}
$
coming from the flux through first $S^3$ only.
The difference is just
$$
\Delta V = V_i - V_f \approx  2|N_2| \,{|r|\over g_s}= {2|N_2|\over g_{YM}^2}
$$
We see that both the domain wall tension $and$ the difference in vacuum
energies before and after are equal to what we found in the open string picture,
albeit rescaled by the amount of brane charge which dissapears!
This is, of course, not surprising because the domain wall relevant in the
closed string case is the same one as that relevant in the open string case, just before the 
large N transition.

We can now put everything together to compute the
action of the instanton corresponding to nucleating a bubble of true
vacuum:
$$
S_I \approx {\pi\over 48} {|N_2| \over g_s}\, {\left| (g \Delta^3)^4\right| \over |r|^3}.
$$
This leads to the decay rate
$$
 \Gamma \sim \exp(-S_I).
$$
Note that the relevant instantons come from D-branes wrapping cycles, 
and thus it is natural for the instanton action to depend on 
$g_s$ in the way the D-brane action does.

This exactly reproduces the results one would have expected from our discussion in section 3. There are two effects controling the stability of the system.
Namely, one is the height of the
potential barrier, which is controlled by here by $g\Delta^3$ and which 
enters the action of the domain wall.
The branes
and anti-branes become more separated the larger the $\Delta$,
and their interactions weaker. In addition, increasing the height of the
potential barrier by making $\Delta$ larger, overcomes the
electrostatic attraction of the branes. 
The other relevant parameter is the relative difference in energies of the 
true and the false 
vacuum, which depends on the ratio of brane to anti-brane
numbers $|N_2/N_1|$. When this is very small, the vacuum with both the branes and the anti-branes is nearly degenerate with the vacuum with just the branes,
and correspondingly, the decay time gets longer as $\Delta V/V$ gets smaller.

\newsec{Open Questions}

In this paper, we have seen how a large $N$ system of branes which
geometrically realizes metastability by wrapping cycles of a non-compact
CY can be dual to a flux compactification which breaks supersysmmetry.
We provided evidence for this duality by studying the limit where the
cycles of the Calabi-Yau are far separated.

There are a number of open questions which remain to be studied:
It would be nice to write down a non-supersymmetric gauge theory
with finite number of degrees of freedom which describes this brane
configuration. It is not clear that this should be possible, but it
would be interesting to settle this question conclusively one way or
the other.  This is also important for other applications in string theory
where dynamics of brane/anti-brane systems is relevant.

Another issue which would be important to study is the phase structure
of our system.  Since we have broken supersymmetry, we have no
holomorphic control over the geometry of the solutions.  It would
be interesting to see what replaces this, and how one should
think of the global phase structure of these non-supersymmetric solutions.
Since the full potential is characterized by the ${\cal N}=2$ prepotential
and the flux data, which in turn are characterized by integrable
data (matrix model integrals of the superpotential), one still expects
that the critical points of the potential should be characterized
in a nice way.  Understanding this structure would be very important.
For example, it would tell us what happens when the separation of the wrapped cycles
is small where one expects to lose metastability due to the open
string tachyons stretching between branes and anti-branes.

One other issue involves the $1/N$ corrections to the holographic
dual theory.  In this paper we have focused on the leading
order in the $1/N$ expansion, i.e. at the string tree level.
How about subleading corrections?
This is likely to be difficult to address, as one expects
non-supersymmetric corrections to string tree level to be
difficult to compute.  The fact that we have a metastable system
suggests that we would not want to push this question
to an exact computation,
as we know the system is ultimately going to decay to a stable
system.  However, we would potentially be interested in studying
the regions of phase space where metastability is just being lost, 
such as when the cycles are close to one another.

Finally, perhaps the most important question is about the embedding of
our mechanism for inducing metastable flux compactification vacua into a 
{\it compact} Calabi-Yau geometry (see \ref\banks{T.~Banks, M.~Dine and L.~Motl,
  ``On anthropic solutions of the cosmological constant problem,''
  JHEP {\bf 0101}, 031 (2001)
  [arXiv:hep-th/0007206].}). At first sight,
it may be unclear whether there would be any obstacles to a compact
embedding of our story.  There is at least an example
similar to what we are expecting in the compact case.  In
\ref\kawa{O.~DeWolfe, A.~Giryavets, S.~Kachru and W.~Taylor,
  ``Type IIA moduli stabilization,''
  JHEP {\bf 0507}, 066 (2005)
  [arXiv:hep-th/0505160].
  }, among the solutions studied,
it was found that flipping the signs of some fluxes leads to apparently
metastable non-supersymmetric vacua.  It would be interesting
to see if this connects to the mechanism introduced in this paper, where
the existence of metastability for flux vacua is a priori expected.
Moreover, from our results it is natural to expect that most flux
vacua {\it do} have metastable non-supersymmetric solutions.  For instance,
in the non-compact case we have studied, only $2$ out of $2^n$ choice of signs
for the $n$ fluxes $2$ were non-supersymmetric.  This suggests that in the study
of flux compactifications, the most natural way to break supersymmetry
is simply studying metastable vacua of that theory, without the necessity of introducing
any additional anti-branes into the system.  It would be very interesting
to study this further.
\bigskip

\centerline{\bf Acknowledgments}
We would like to thank N. Arkani-Hamed,
K. Becker, M. Becker, R. Bousso, E. Imeroni,
K. Intriligator, K. Kim, J. McGreevy, L. Motl, D. Sahakyan, D. Shih, W. Taylor and B. Zwiebach
for valuable
discussions.  We would also like to thank the Stony Brook Physics department and the
fourth Simons Workshop in Mathematics and Physics for their hospitality while this project was 
initiated.

The research of M.A. and C.B. is supported in part by the UC Berkeley Center for
Theoretical Physics.
The research of M.A. is also supported by a DOI OJI Award, the Alfred P. Sloan Fellowship and the
NSF grant PHY-0457317.
The research of J.S. and C.V. is supported in part by NSF grants PHY-0244821 and DMS-0244464. The research of J.S. is also supported in part by the Korea Foundation for Advanced Studies.

\listrefs

\end

At  low energy theory is described by a

This theory to have two different descriptions, depending on the scale.
At energies
$$
{\Lambda_0}\geq \Delta
$$
the theory is well described by the ${\cal N}=2$
$U(N_1+N_2)$ gauge theory with supersymmetry broken by the superpotential $W(\Phi)$.
At low energies
$$
{\Lambda_0} \leq \Delta
$$
the natural description is in terms of an effective $pure$
${\cal N}=1$ $U(N_1)\times U(N_2)$ gauge theory, perturbed by some
high dimension operators.
Certain quantities in the theory do not depend on the scale ${\Lambda_0}$,
so we can compute them directly either at high energies, or at low energies.
One such quantity is the effective superpotential. Others, like masses of the fields,
do depend on the D-terms, and thus depend on the energy scale.
Since the large N dual theory is abelian, with ${\cal N}=2$ space time supersymmetry,
we expect to be able to compute reliably both the D-terms and the F-terms from the geometry.

In the limit where all the $a_i$'s are widely separated, the systems
are essentially non-interacting to be much smaller than the
separations between them, the sizes of the $S^3$'s were dual to the
vevs of the gaugino condensates of the corresponding gauge groups,
i.e, if we denote by $S_k = {1\over 32 \pi^2} Tr W^k_\alpha
W^{k\alpha}$, then

In this case, the open
string theory side has a simple gauge theory description at low
energies, which corresponds to starting with $U(N)$, ${\cal N}=2$
supersymmetric Yang-Mills theory where $N=\sum_{=1}^n N_n$, with
supersymmetry broken to ${\cal N}=1$ by a superpotential ${W}(\Phi)$
for the adjoint scalar. The classical vacua of this, correspond to
critical points of the superpotential, and then we have a choice of
how many eigenvalues $N_k$ we want to put in the $k$'th vacuum.
The vacua with different choices how many branes we put on each cycle
are separated by energy barriers. To climb over these, the branes have to get more massive first.

As before, the sizes of the $S^3$'s are related to
now wrapping $N_1$ D5
branes on the While they are all in the same homology class, there is
an energy

This theory
had a simple gauge theory description, below the string scale, which
corresponds to starting with $U(N)$, ${\cal N}=2$ supersymmetric
Yang-Mills theory where $N=\sum_{k} N_k$, with supersymmetry broken to
${\cal N}=1$ by a superpotential ${W}(\Phi)$ for the adjoint scalar.
The theory has different vacya
depending on how many branes we put on each ${\bf P}^1$.
With $N_k$ branes on the $k$-th ${\bf P}^1$, at low energies, below the scale set by the superpotential, the theory is a
pure ${\cal N}=1$ $\prod_{k}U(N_k)$ gauge theory, which confines.
The vacua with different
choices of how many branes we put on each cycle are separated by energy
barriers. To climb over these, the branes have to get more massive
first.

Note that, to this order, the superpotential can be written as
$$
{\cal W}(S) = \sum_{i=1,2} \alpha_i(\Lambda_0) S_i  +
N_i {S_i\over 2 \pi i}(\log({S_i\over \Lambda_0^3}) - 1)
$$
where
$$
2 \pi i \alpha_i(\Lambda_0)= -\log({\Lambda_i \over \Lambda_0})^{3 N_i}, \qquad i=1,2
$$
and
\eqn\scalematch{
({\Lambda_1\over g \Delta^3})^{N_1} = ({\Lambda \over \Lambda_0})^{2(N_1+N_2)}=
({\Lambda_2\over g \Delta^3})^{N_2}.
}
The is is the apropriate form of the superpotential at scales below
$\Delta$, where the two theories decouple.  From the open string
perspective, the bifundamental strings connecting the two stacks of
branes have masses of order $\Delta$, and at energies below this, it
is natural to integrate them out.  The running coupling constants
$\alpha_{1,2}(\Lambda_0)$ of the two low energy theories agree with
$\alpha(\Lambda_0)$ at scale $\Delta$, when the strong coupling scales
satisfy \scalematch .

